\documentclass[Journal]{IEEEtran}
\IEEEoverridecommandlockouts
\usepackage{cite}
\usepackage{amsmath,amssymb,amsfonts}
\usepackage{algorithmic}
\usepackage{graphicx,graphics}
\usepackage{textcomp}
\usepackage{xcolor}
\usepackage{algorithm}
\usepackage{multirow}
\usepackage[acronym,nonumberlist,nopostdot]{glossaries}
\usepackage{caption}
\usepackage{subcaption}
\usepackage{orcidlink}

\title{XR Capacity Enhancement through Multi-Connected XR Tethering Groups \\
}

\makeglossaries
\newacronym{5g}{5G}{5th generation}
\newacronym{5ga}{5G-A}{5th generation-advanced}
\newacronym{3gpp}{3GPP}{3rd Generation Partnership Project}
\newacronym{ack}{ACK}{acknowledgment}
\newacronym{ar}{AR}{augmented reality}
\newacronym{bler}{BLER}{block error rate}
\newacronym{cb}{CB}{code block}
\newacronym{cbg}{CBG}{code block group}
\newacronym{cc}{CC}{Chase combining}
\newacronym{csi}{CSI}{channel state information}
\newacronym{cqi}{CQI}{channel quality indicator}
\newacronym{du}{DU}{Dense Urban}
\newacronym{dl}{DL}{downlink}
\newacronym{df}{DF}{decode-and-forward}
\newacronym{embb}{eMBB}{enhanced mobile broadband}
\newacronym{ecdf}{eCDF}{empirical cumulative distribution function}
\newacronym{gNodeB}{gNodeB}{Next-Generation Node B}
\newacronym{harq}{HARQ}{hybrid automatic repeat request}
\newacronym{illa}{ILLA}{Inner Loop Link Adaptation}
\newacronym{inh}{InH}{Indoor Hotspot}
\newacronym{ici}{ICI}{inter-cell interference}
\newacronym{la}{LA}{Link Adaptation}
\newacronym{los}{LOS}{line-of-sight}
\newacronym{llr}{LLR}{log-likelihood ratio}
\newacronym{kpi}{KPI}{key performance indicator}
\newacronym{mmib}{MMIB}{mean mutual information per bit}
\newacronym{mmwave}{mmWave}{millimeter-wave}
\newacronym{mimo}{MIMO}{Multiple-Input and Multiple-Output}
\newacronym{mbs}{MBS}{multicast and broadcast services}
\newacronym{mmseirc}{MMSE-IRC}{minimum mean square error-interference rejection combining}
\newacronym{mcs}{MCS}{modulation and coding scheme}
\newacronym{mr}{MR}{mixed reality}
\newacronym{nr}{NR}{new radio}
\newacronym{nack}{NACK}{negative acknowledgment}
\newacronym{nlos}{NLOS}{non-line-of-sight}
\newacronym{ofdm}{OFDM}{orthogonal frequency-division multiplexing}
\newacronym{olla}{OLLA}{Outer Loop Link Adaptation}
\newacronym{pdb}{PDB}{packet delay budget}
\newacronym{ptm}{PTM}{point-to-multipoint}
\newacronym{ptp}{PTP}{point-to-point}
\newacronym{prb}{PRB}{Physical Resource Block}
\newacronym{pdsch}{PDSCH}{physical downlink shared channel}
\newacronym{pdcch}{PDCCH}{physical downlink control channel}
\newacronym{qos}{QoS}{quality of service}
\newacronym{rrm}{RRM}{radio resource management}
\newacronym{rx}{Rx}{Receive}
\newacronym{sc}{SC}{selection combining}
\newacronym{scs}{SCS}{Selection Combining Scheme}
\newacronym{sscs}{SSCS}{Selection/Soft Combining Scheme}
\newacronym{svd}{SVD}{singular value decomposition}
\newacronym{softdf}{soft-DF}{soft decode-and-forward}
\newacronym{sinr}{SINR}{signal-to-interference-plus-noise ratio}
\newacronym{sls}{SLS}{system-level simulator}
\newacronym{tbler}{TBLER}{transport block error rate}
\newacronym{thgr}{TGr}{tethering group}
\newacronym{thgrs}{TGrs}{tethering groups}
\newacronym{tb}{TB}{transport block}
\newacronym{tx}{Tx}{Transmit}
\newacronym{tdd}{TDD}{Time Division Duplex}
\newacronym{ue}{UE}{User Equipment}
\newacronym{ue-x}{UE-X}{User Equipment-extended reality}
\newacronym{ue-t}{UE-T}{User Equipment with Tethering}
\newacronym{ul}{UL}{uplink}
\newacronym{urllc}{URLLC}{Ultra Reliable Low Latency Communications}
\newacronym{vr}{VR}{virtual reality}
\newacronym{wlan}{WLAN}{wireless local-area network}
\newacronym{xr}{XR}{Extended Reality}

\glsaddall

\author{Muhammad Ahsen \orcidlink{0009-0005-2275-3091}, \IEEEmembership{Graduate Student Member, IEEE}, Boyan Yanakiev, Claudio Rosa, and Ramoni Adeogun \orcidlink{0000-0003-1118-7141}, \IEEEmembership{Senior Member, IEEE}

\thanks{Muhammad Ahsen and Ramoni Adeogun are with the Department of Electronic Systems, Technical Faculty of IT and Design, Aalborg University, 9220 Aalborg, Denmark (e-mail: muah@es.aau.dk; ra@es.aau.dk).} 
\thanks{Boyan Yanakiev and Claudio Rosa are with the Nokia, DK-9220 Aalborg, Denmark (e-mail:boyan.yanakiev@nokia.com; claudio.rosa@nokia.com)}}

\def\BibTeX{{\rm B\kern-.05em{\sc i\kern-.025em b}\kern-.08em
    T\kern-.1667em\lower.7ex\hbox{E}\kern-.125emX}}
\def\and{%
  \end{tabular}%
  \hskip .5em \@plus.17fil\relax
  \begin{tabular}[t]{c}}

\begin{document}
\newcommand{\roa}[1]{\textcolor{blue}{[Ramoni: #1]}}

\maketitle

\begin{abstract}
\acrfull{xr} applications have limited capacity in \acrfull{5ga} cellular networks due to high throughput requirements coupled with strict latency and high reliability constraints. 
To enhance \acrshort{xr} capacity in the \acrfull{dl}, this paper investigates multi-connected \acrshort{xr} \acrfull{thgrs}, comprising an \acrshort{xr} device and a cooperating \acrshort{5ga} device.
This paper presents investigations for two types of cooperation within \acrshort{xr} \acrshort{thgr}, i.e., \acrfull{sc} and soft combining and their impact on the \acrshort{xr} capacity of the network.
These investigations consider joint \acrfull{harq} feedback processing algorithm and also propose enhanced joint \acrfull{olla} algorithm to leverage the benefits of multi-connectivity.
These enhancements aim to improve the spectral efficiency of the network by limiting \acrshort{harq} retransmissions and enabling the use of higher \acrfull{mcs} indices for given \acrfull{sinr}, all while maintaining or operating below than the target \acrfull{bler}.
Dynamic system-level simulation 
demonstrate that \acrshort{xr} \acrshort{thgr}s with soft combining achieve performance improvements of 23 - 42\% in \acrshort{xr} capacity with only \acrshort{xr} users 
and 38-173\% in the coexistence scenarios consisting of \acrshort{xr} users and \acrfull{embb} user.
Furthermore, the enhanced joint \acrshort{olla} algorithm enables similar performance gains even when only one device per \acrshort{xr} \acrshort{thgr} provides \acrfull{csi} reports, compared to scenarios where both devices report \acrshort{csi}.
Notably, \acrshort{xr} \acrshort{thgr}s with soft combining also enhance \acrshort{embb} throughput in coexistence scenarios.
\end{abstract}

\begin{IEEEkeywords}
\acrshort{xr}, Wireless tethering, Cooperative Communication, \acrshort{mbs}, \acrshort{harq}, \acrlong{thgr}.
\end{IEEEkeywords}

\section{Introduction}

\gls{xr} 
has applications across a wide range of domains including education, healthcare, gaming and autonomous driving etc.
It is recognized as one of the key use cases for future wireless networks \cite{bib1}.
\gls{xr} services typically impose stringent \gls{qos} requirements, with latency constraints ranging from 5 to 15 milliseconds and data rate demands between 30 and 60 Mbps \cite{bib2}.  
Additionally, high reliability typically at 99\% level is essential. 

Several studies have investigated the feasibility of \gls{5ga} cellular networks for supporting \gls{xr} applications \cite{bib3-1,bib3-2, bib3-7, bibr4}.  
However, due to the strict \gls{qos} requirements, the number of \gls{xr} users that \gls{5ga} can support remains limited. 
This limitation underscores the need for more efficient wireless network solutions capable of delivering seamless \gls{xr} experience at scale.

Among various techniques under investigation, multi-connectivity has emerged as one of the promising approach for enhancing \gls{xr} performance \cite{bib7,bib8, bib4-2, bib4-1}. 
Research on \acrlong{mmwave} networks \cite{bib7} and \cite{bib8} has shown that multi-connectivity can improve user density and mitigate the effects of blockages, thereby enhancing the \gls{xr} user experience.
In \cite{bib4-2}, a dual connectivity architecture leveraging both sub-6 GHz and \acrlong{mmwave} frequencies, combined with mobile edge capability, was proposed to improve \gls{vr} reliability.
Further, \cite{bib4-1} examined \gls{xr} capacity improvements in coexistence scenarios involving \gls{xr} and broadband traffic by employing multi-connectivity across sub-6 GHz and \acrlong{mmwave} bands.
Despite these advantages, traditional multi-connectivity approaches, which rely on multiple \gls{ptp} connections, consume significant radio resources. 
Similar approaches have been studied for \gls{urllc} applications in the sub-6 GHz band. 
Although these approaches improve transmission reliability, they degrade spectral efficiency amd overall network throughput \cite{bib9,bib10}. 
Thus direct application of such \gls{ptp}-based multi-connectivity methods to \gls{xr} over sub-6 GHz frequencies is impractical, as it further constrains \gls{xr} capacity.

To address this, we explore \gls{xr} \gls{thgrs} as a means of utilizing multi-connectivity benefits without the associated cellular network resource overhead. 
\gls{xr} \gls{thgr}s enable multiple connections (direct and indirect) for an \gls{xr} device via \gls{ptm} transmissions \cite{bib2-our,bib2-our2, bib2-our3}.
Each \gls{thgr} consists of an \gls{xr} device and a \gls{5ga} device with tethering capability referred to as the tethering device, which are also interconnected via a tethering link \cite{bib6}.
Within a \gls{thgr}, the tethering device forwards received data to the \gls{xr} device, creating cooperative communication between the two devices.

We investigate two cooperation schemes for \gls{thgr}s under \gls{ptm} transmission from the cellular network, i.e., \gls{sc} and soft combining.
In \gls{sc}, the \gls{xr} device can select the \gls{tb} from either the direct link or the indirect link.
In soft combining, the \gls{xr} device performs soft-bit level combining of \gls{tb}s received from both links.
This is analogous to \gls{harq} with \gls{cc} where soft bits from repeated transmissions are combined to improve decoding performance\cite{bib2-3} \cite{bib2-18}. 
In case of \gls{harq} retransmission with \gls{cc}, cellular network's radio resources are doubled for sending the same \gls{tb} assuming the \gls{harq} retransmission is scheduled with same \gls{mcs}.
However, in \gls{xr} \gls{thgr}, \gls{xr} device has two copies of the same \gls{tb} (or two sets of the soft bits corresponding to the same \gls{tb}) which traversed through different channel conditions 
while utilizing single cellular network resources (assuming tethering link operate on different carrier frequency). 

Prior work \cite{bib2-our} and \cite{bib2-our2} has proposed joint \gls{harq} feedback processing algorithm for \gls{ptm} transmission towards the \gls{xr} \gls{thgr} to enhance the spectrum efficiency by incorporating the cooperation benefits in \gls{harq} retransmission from the network. 
However, these studies assumed a static \gls{mcs} and did not consider dynamic \gls{la} with investigations limited to link-level.
In \cite{bib2-our3}, we extended this investigation by evaluating \gls{xr} \gls{thgr}s using \gls{sc} with dynamic \gls{la} in system-level simulations.
This paper further advances our work by considering soft combining with dynamic \gls{la} for \gls{xr} \gls{thgr}s in system-level evaluations and its impact on \gls{xr} capacity.
Our focus is to assess how \gls{xr} \gls{thgr}s employing selection/soft combining affect overall \gls{xr} capacity of the network under \gls{ptm} transmission.

\gls{la} is employed to select the optimal \gls{mcs} based on \gls{cqi} indicator provided in the \gls{csi} report by the \gls{ue} \cite{bib3-3}.
To address imperfections in \gls{cqi} indicator, stemming from quantization errors, measurement inaccuracies, feedback delays and temporal channel fluctuations, \gls{olla} is applied.
\gls{olla} dynamically adjusts the \gls{mcs} selection using \gls{harq} feedback to maintain a predefined target \gls{bler} \cite{bib3-4}, thereby improving the robustness and reliability of the \gls{la} mechanism.

The investigations in \cite{bib2-our3} concludes that traditional \gls{olla}, which relies solely on \gls{harq} feedback from a single \gls{ue} (i.e., \gls{olla} fixed to one \gls{ue}) is insufficient to translate \gls{thgr} cooperative gains into \gls{la} improvements.
Instead, a joint \gls{olla} is required to leverage these cooperative benefits in \gls{la}. 
However, the joint \gls{olla} algorithm proposed in \cite{bib2-our3} for \gls{ptm} transmission with cooperating users is limited to \gls{sc} use cases only.
In this paper, we are extending our previous work of \cite{bib2-our3}, by proposing an enhanced joint \gls{olla} algorithm for \gls{ptm} transmission with cooperating users that supports both \gls{sc} and soft combining.
The goal is to exploit device cooperation within the \gls{thgr} to enable higher \gls{mcs} selection, reducing latency and improving spectral efficiency.
This can increase the number of \gls{xr} users meeting \gls{qos} requirements and/or free up resources for other services.

Coexistence scenarios, i.e., \gls{xr} users with \gls{embb} users have also been examined in available literature related to \acrshort{xr} \cite{bib3-5,bib3-6}, which showed that \gls{embb} users negatively impact \gls{xr} capacity due to \gls{ici}.
However, these studies considered only single-link-connected \gls{xr} users.
Here, we investigate \gls{xr} \gls{thgr}s in the presence of \gls{embb} users to assess their impact on \gls{xr} performance or capacity and potential improvements in \gls{embb} throughput, given the lower resource consumption of \gls{xr} \gls{thgr}s.

Additionally, we extend the limited \gls{cb} based soft combining initially proposed in \cite{bib2-our2}, which was confined to a single \gls{thgr} and static \gls{la}.
Our current investigation evaluates \gls{cb}-based soft combining in a multi-cell network with multiple \gls{xr} \gls{thgr}s and dynamic \gls{la}, measuring its impact on \gls{xr} capacity.
In this approach, the \gls{xr} device treats \gls{cb}s received from the tethering device similar to \gls{cbg} based \gls{harq} retransmissions \cite{bib3-1} but without consuming additional radio resources (assuming tethering link operates on a separate carrier).

Our major contributions are summarized as follows
\begin{itemize}
    \item We investigate the \gls{xr} capacity of the network with \gls{xr} \gls{thgr}s with \gls{sscs} under \gls{ptm} transmission, employing joint \gls{harq} and \gls{la} in a system-level simulation calibrated according to \cite{bib2} for \gls{du} and \gls{inh} scenarios, each with unique channel conditions, cell deployments amd user densities \cite{bib13}.
    \item We propose an enhanced joint \gls{olla} algorithm for \gls{ptm} transmission that supports both \gls{sc} oand \gls{sscs}, leveraging cooperation to enable higher \gls{mcs} selection.
    \item We analyze the performance of \gls{xr} \gls{thgr}s with \gls{sscs} in the presence of \gls{embb} user, evaluating both the impact on \gls{xr} \gls{thgr} performance and the potential throughput gains for \gls{embb}.
    \item We investigate \gls{xr} capacity enhancements through limited \gls{cb}-based cooperation within a \gls{thgr} using \gls{sscs}.
\end{itemize}

The structure of the article is as follows: Section \ref{systemmodel} outlines the system model followed by the cooperation and adaptation techniques for \gls{xr} \gls{thgr} in Section \ref{cooptech} and \gls{cb}-based cooperation in Section \ref{cbcoop}.
Section \ref{evam} details the evaluation methodology.
Simulation results are discussed in Section \ref{simrst} and the conclusion is presented in \ref{conc}.

\section{System Model}
\label{systemmodel}
\begin{figure}
\centerline{\includegraphics[width=7cm]{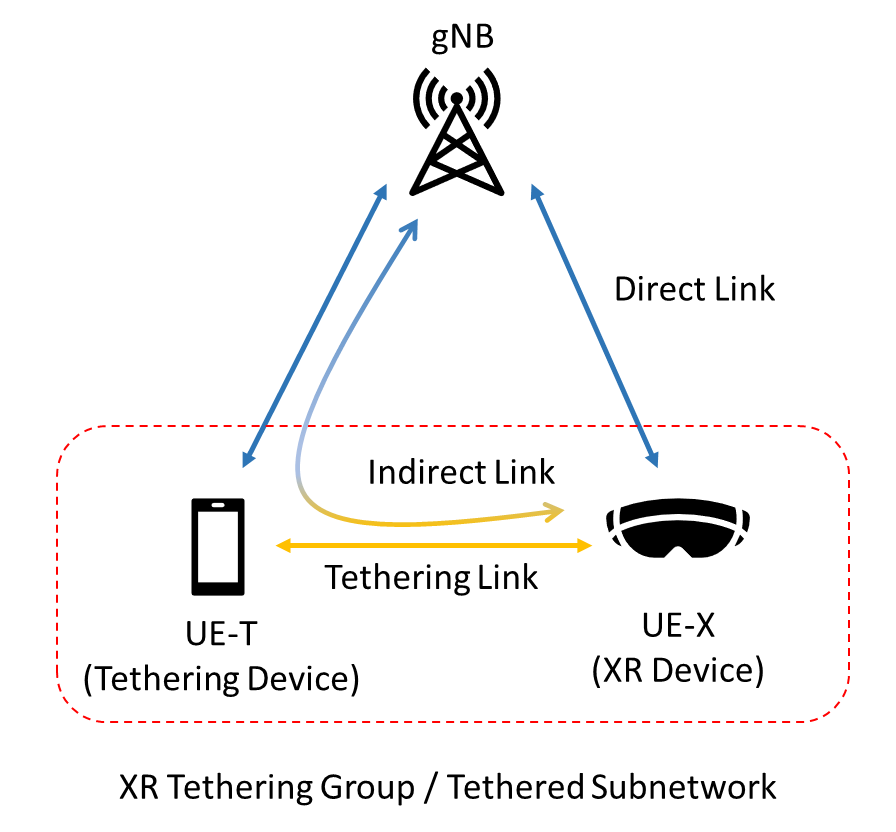}}
\caption{\gls{xr} \gls{thgr}.}
\label{fig:TGr}
\end{figure}

This study considers a multi-cell \gls{dl} network comprising multiple \gls{xr} \gls{thgr}s.
Each \gls{thgr} consists of an \gls{xr} \gls{ue} (\acrshort{ue-x}) and a tethering \gls{ue} (\acrshort{ue-t}).
Both \gls{ue}s in a \gls{thgr} are connected to the same \gls{gNodeB} and are members of the same \gls{5ga} multicast session \cite{bib11}.
However, each \gls{thgr} is associated with a distinct multicast session to avoid inter-group overlap.
The location of \gls{thgr}s within a cell is random.
Within a \gls{thgr}, \acrshort{ue-t} is positioned at a fixed distance $x$ from \acrshort{ue-x}. 
Each \gls{5ga} multicast session delivers identical data to both \gls{ue}s in a \gls{thgr} using \gls{ptm} transmission, applicable to both initial transmissions and any \gls{harq} retransmissions.
In the remainder of the paper, a multicast session with cooperating member \gls{ue}s is referred to as a cooperative multicast session.
Within a \gls{thgr}, \acrshort{ue-t} assists \acrshort{ue-x} by relaying \gls{xr} data over a tethering link \cite{bib6}, effectively providing \acrshort{ue-x} with an additional indirect link from the \gls{gNodeB} (via \acrshort{ue-t}), as shown in Figure \ref{fig:TGr}.

Asynchronous \gls{harq} is modeled, where each \gls{ue} within a \gls{thgr} independently provides \gls{harq} feedback based on its initial decoding of the \gls{tb} \cite{bib12}.
We consider both \gls{illa} and \gls{olla} for \gls{ptm} transmissions configured for \gls{xr} \gls{thgr}s.
While \gls{illa} governs the selection of \gls{mcs} based on \gls{csi}, \gls{olla} adjust \gls{mcs} to achieve the target \gls{bler}.

The tethering link is assumed to operate on a carrier frequency that is different from the one used by the cellular network.
It may utilize either \gls{wlan} or \gls{5g} sidelink technologies \cite{bib6}.
While the performance of the tethering link impacts the overall performance of the \gls{thgr}, this study focuses on assessing the upper-bound capacity gains achievable through multi-connected \gls{thgr}.
In line with the objective, the tethering link is modeled with the following assumptions:
\begin{itemize}
    \item For the direct link (\gls{ue}-\gls{gNodeB}) and the tethered link (\acrshort{ue-x} to \acrshort{ue-t}), we assume the usage of distinct radio interfaces running on distinct carrier frequencies to enable simultaneous communication with no notable in-device-coexistence issues.
    \item The tethering link is additionally assumed to have enough capacity to handle the data load, ideal channel conditions and zero processing/decoding and forwarding delays.
\end{itemize}

\section{Cooperation and Adaptation Techniques for \gls{xr} \gls{thgr}}
\label{cooptech}
\subsection{Cooperation in \gls{xr} \gls{thgr}}
We consider following two cooperation schemes for \gls{thgr}.
\begin{itemize}
    \item \gls{scs}: In this scheme, \acrshort{ue-t} decodes the received \gls{tb} and forwards it to \acrshort{ue-x} using the tethering link, following \gls{df} algorithm \cite{bib18-2}. \acrshort{ue-x} performs \gls{sc} over the two \gls{tb} copies, one from the direct \gls{gNodeB}-\acrshort{ue-x} link and one from the indirect \gls{gNodeB}-\acrshort{ue-t}-\acrshort{ue-x} link. If \acrshort{ue-x} successfully decodes the \gls{tb} via the direct link, it discards the \gls{tb} if received from \acrshort{ue-t}. If decoding fails at \acrshort{ue-x}, then it uses the \gls{tb} if received from \acrshort{ue-t} via a tethering link.   
    \item \gls{sscs}: This scheme extends \gls{scs} by introducing soft combining, depending on decoding outcomes at both \gls{ue}s of the \gls{thgr}.
    If \acrshort{ue-t} successfully decodes the \gls{tb}, it forwards the decoded \gls{tb} to \acrshort{ue-x} using the \gls{df} algorithm.
    If \acrshort{ue-t} fails to decode the \gls{pdsch} but successfully decodes the \gls{pdcch}, it forwards soft values corresponding to that \gls{tb}  using a \gls{softdf} approach \cite{bib2-1}. \acrshort{ue-x} then applies either \gls{sc} or soft combining, based on its own decoding outcome of \gls{pdsch} and \gls{pdcch}. The purpose of \gls{sscs} is to reduce the data overhead on the tethering link and/or minimize the computational burden on \acrshort{ue-x}, by switching intelligently between \gls{sc} and soft combining. The details regarding the overhead impact of different cooperation scheme can be found in our prior work in \cite{bib2-our2}. In soft combining, \acrshort{ue-x} adds the soft bits (\gls{llr}) from the two sources, direct link and indirect link. The sign and magnitude of each \gls{llr} represents the bit decision and its confidence level, respectively \cite{bib2-2}.    
\end{itemize}

The behavior of the \gls{thgr} under different decoding scenarios of \gls{pdcch} and \gls{pdsch} assuming Rank-1 transmission is summarized in Table \ref{tab:tgrworking}. 
In the table, "D" indicates successful decoding, while "N/D" indicates decoding failure. 
\gls{tb}\textsubscript{T} and \gls{tb}\textsubscript{X} represent the decoded \gls{tb}s at \acrshort{ue-t} and \acrshort{ue-x}, respectively. 
Similarly soft\gls{tb}\textsubscript{T} and soft\gls{tb}\textsubscript{X} represent soft bits of the \gls{tb} at \acrshort{ue-t} and \acrshort{ue-x} respectively.
From Table \ref{tab:tgrworking}, it can be seen that \gls{scs} offers gains in scenarios 2 and 3 only, whereas \gls{sscs} offers gains in scenario 5 as well if \gls{tb} is decoded after soft combining, in addition to scenarios 2 and 3. 
To fully exploit the potential performance improvements offered by these cooperation schemes, modifications are required in both the \gls{harq} retransmission logic and the \gls{la} algorithms at the network side.
These updates are essential to adaptively select appropriate \gls{mcs} values and retransmission policies based on decoding pattern after cooperation between the \gls{ue}s of the \gls{thgr}.

\begin{table*}
\caption{\gls{pdcch} and \gls{pdsch} Decoding Status (Successfully Decoded: D and Fail to Decode: N/D) and Cooperation Strategies for the Tethering Group with \gls{scs} and \gls{sscs}}
\begin{center}
\renewcommand{\arraystretch}{1.2}
\begin{tabular}{|c|c|c|c|c|c|c|c|c|c|}
\hline
 \textbf{Scenario} & \multicolumn{2}{|c|}{\textbf{\acrshort{ue-t}}}& \multicolumn{2}{|c|}{\textbf{\acrshort{ue-x}}} & \multicolumn{2}{|c|}{\textbf{\gls{scs}}} & \multicolumn{2}{|c|}{\textbf{\gls{sscs}}} \\
\hline
& \multirow{2}{*}{\textbf{\gls{pdcch}}}& \multirow{2}{*}{\textbf{\gls{pdsch}}} & \multirow{2}{*}{\textbf{\gls{pdcch}}}& \multirow{2}{*}{\textbf{\gls{pdsch}}} & \textbf{Relaying} & \textbf{Combining} & \textbf{Relaying} & \textbf{Combining}\\
& &  & &  & at \acrshort{ue-t} & at \acrshort{ue-x} & at \acrshort{ue-t} & at \acrshort{ue-x}\\
\hline

1 & D & D & D & D & \multirow{3}{*}{\gls{tb}\textsubscript{T}} & Discard \gls{tb}\textsubscript{T} & \multirow{3}{*}{\gls{tb}\textsubscript{T}} & Discard \gls{tb}\textsubscript{T}\\
\cline{1-5}
\cline{7-7}
\cline{9-9}
2 & D & D & D & N/D &  & \multirow{2}{*}{Use \gls{tb}\textsubscript{T}} &  & \multirow{2}{*}{Use \gls{tb}\textsubscript{T}}\\
\cline{1-5}

3 & D & D & N/D & N/D &  &  &  & \\
\cline{1-9}
4 & D & N/D & D & D & \multirow{6}{*}{-} & \multirow{6}{*}{-} & \multirow{3}{*}{soft\gls{tb}\textsubscript{T}} & Discard soft\gls{tb}\textsubscript{T} \\
\cline{1-5}
\cline{9-9}
5 & D & N/D & D & N/D &  &  &  & soft\gls{tb}\textsubscript{T} $+$ soft\gls{tb}\textsubscript{X}\\
\cline{1-5}
\cline{9-9}
6 & D & N/D & N/D & N/D &  &  &  & Discard soft\gls{tb}\textsubscript{T}\\
\cline{1-5}
\cline{8-9}
7 & N/D & N/D & D & D &  &  & \multirow{3}{*}{-} & \multirow{3}{*}{-}\\
\cline{1-5}
8 & N/D & N/D & D & N/D &  &  &  &  \\
\cline{1-5}
9 & N/D & N/D & N/D & N/D &  &  &  &  \\
\hline
\end{tabular}
\label{tab:tgrworking}
\end{center}
\end{table*}

\subsection{\gls{harq} in \gls{xr} \gls{thgr}}
For each \gls{thgr}, a joint \gls{harq} feedback processing algorithm proposed in \cite{bib2-our2} is implemented at the \gls{gNodeB}.
In case of \gls{sscs}, \acrshort{ue-x} provides additional \gls{harq} feedback only if it performs soft combining whereas in \gls{scs}, \acrshort{ue-x} does not provide any additional \gls{harq} feedback after performing \gls{sc}.
The joint \gls{harq} feedback processing algorithm limits retransmission to only the following scenarios 
\begin{itemize}
    \item For \gls{scs}: \gls{harq} retransmission is triggered only if both \acrshort{ue-t} and \acrshort{ue-x} fail to decode the \gls{tb}.
    \item For \gls{sscs}: \gls{harq} retransmission is triggered only if \acrshort{ue-x} fails to decode the \gls{tb} even after soft combining the soft bits received from direct and indirect link.
\end{itemize}

This joint \gls{harq} feedback processing algorithm increases spectral efficiency by reducing unnecessary retransmissions from the \gls{gNodeB}, thereby exploiting the benefits of intra-\gls{thgr} cooperation. 

\subsection{\gls{la} in \gls{xr} \gls{thgr}}

\subsubsection{\gls{illa} in \gls{xr} \gls{thgr}}
We study \gls{mcs} selection under two \gls{csi} reporting configurations per \gls{xr} \gls{thgr}:
\begin{itemize}
    \item \gls{csi} \acrshort{ue-x}: Only \acrshort{ue-x} provides periodic \gls{csi} reports. The \gls{gNodeB} selects the \gls{mcs} and precoding weights for the \gls{ptm} transmission based solely on the \gls{cqi} and channel estimates from \acrshort{ue-x}.
    \item \gls{csi} Best: Both \acrshort{ue-t} and \acrshort{ue-x} per \gls{thgr} provide periodic \gls{csi} reports. The \gls{gNodeB} then selects the better of the two \gls{csi} reports (based on higher \gls{cqi}) to determine the \gls{mcs} and precoding weights for the \gls{ptm} transmission towards that \gls{xr} \gls{thgr}.
\end{itemize}

\subsubsection{\gls{olla} in \gls{xr} \gls{thgr}}
\label{jointolla}
As per the existing \gls{la} algorithm, the \gls{gNodeB} can select the \gls{mcs} index $\eta^*$ for a transmission as follows
\begin{equation}
    \eta^* = \arg \max_{\eta} \left(R_{\eta}|\text{\acrshort{tbler}\textsubscript{\gls{ue}}} \leq \text{\acrshort{tbler}\textsuperscript{T}}\right)
\end{equation}
where \acrshort{tbler}\textsubscript{\gls{ue}} is the perceived \acrlong{tbler} of the \gls{ue} (could be \acrshort{ue-x} or \acrshort{ue-t}), \text{\acrshort{tbler}\textsuperscript{T}} is the target \acrshort{tbler} and $R_{\eta}$ denotes the data rate associated with \gls{mcs} index $\eta$.
Practically, the \gls{ue} determines its \gls{dl} \gls{mcs} by comparing its measured \gls{sinr} with an internal lookup table mapping \gls{sinr} values to  \acrshort{tbler} for each \gls{mcs}.
This result is then reported to the \gls{gNodeB} in \gls{cqi} report. 
Notably, this step is carried out without considering cooperation, as done in conventional \gls{ue}. 
The reported \gls{sinr} is further adjusted to account for the inaccuracies in the \gls{cqi} by an \gls{olla} offset, based on \gls{harq} feedback, which also does not reflect cooperation gains directly.

Therefore, for \gls{xr} \gls{thgr} we want to select the \gls{mcs} index $\eta^*$ based on the perceived post-cooperation \acrshort{tbler} of \gls{thgr}, i.e., \text{\acrshort{tbler}\textsubscript{\gls{thgr}}}, as follows
\begin{equation}
    \eta^* = \arg \max_{\eta} \left(R_{\eta}|\text{\acrshort{tbler}\textsubscript{\gls{thgr}}} \leq \text{\acrshort{tbler}\textsuperscript{T}}\right)
    \label{eq:eolla}
\end{equation}
However, \text{\acrshort{tbler}\textsubscript{\gls{thgr}}} is not directly available, as both \gls{ue}s (\acrshort{ue-x} and \acrshort{ue-t}) provide \gls{harq} feedback based on its initial decoding of the \gls{tb}. 
In addition to the initial \gls{harq} feedback, \acrshort{ue-x} in \gls{sscs} cooperation scheme also provide 2\textsuperscript{nd} \gls{harq} feedback corresponding to the same \gls{tb} after soft combining (when the initial decoding fails at both \gls{ue} but is successful after combining soft bits at \acrshort{ue-x}).
To implement \eqref{eq:eolla} and fully translate the cooperation benefits in \gls{la}, we propose an enhanced joint \gls{olla} algorithm.
This algorithm modifies the \gls{sinr} used for \gls{mcs} selection by incorporating the cooperation gain and handles multiple \gls{harq} feedback corresponding to the same \gls{tb} for \gls{sscs}. 
The enhanced joint \gls{olla} algorithm for \gls{ptm} mutlicast transmission to \gls{thgr} is summarized in Algorithm \ref{algo:olla1}.
It first computes a joint \gls{harq} feedback ($HF_J$) based on the individual \gls{harq} feedbacks from \acrshort{ue-x} and \acrshort{ue-t}, which is then used for calculating the \gls{olla} correction offset.
An \gls{olla} correction offset $\Delta_{OLLA}$ is applied to the \gls{sinr} to compensate for \gls{cqi} inaccuracies and include the benefits of cooperation.
If $\gamma^{\gls{thgr}_i}$ is the reported \gls{sinr} of \gls{thgr}, $i$, the effective \gls{sinr}, $\gamma_{eff}^{\gls{thgr}_i}$, for \gls{mcs} selection for \gls{ptm} multicast transmission is calculated as 
\begin{equation}
    \gamma_{eff}^{\gls{thgr}_i} = \gamma^{\gls{thgr}_i} - \Delta_{OLLA}(i)
\end{equation}
The offset $\Delta_{OLLA}(i)$ is initialized to a fixed value and is decreased by $\Delta_{down}$ when $HF_J$ is positive \gls{ack} (promoting higher \gls{mcs} selection) and increased by $\Delta_{up}$ when $HF_J$ is \gls{nack} (promoting lower \gls{mcs} selection). 
$\gamma^{\gls{thgr}_i}$ is determined by the configured \gls{csi} reporting scheme and can be calculated as

\begin{equation}
    \gamma^{\gls{thgr}_i} = \begin{cases}
        \max \left( \gamma^{\acrshort{ue-x}_i}, \gamma^{\acrshort{ue-t}_i}\right), & \text{if \gls{csi} Best} \\
        \gamma^{\acrshort{ue-x}_i}, & \text{if \gls{csi} \acrshort{ue-x}} \\
    \end{cases}
\end{equation}
where $\gamma^{\acrshort{ue-x}_i}$ and $ \gamma^{\acrshort{ue-t}_i}$ is the \gls{sinr} reported by \acrshort{ue-x} and \acrshort{ue-t} of \gls{thgr} $i$ in the \gls{csi}.
The enhanced joint \gls{olla} implements \eqref{eq:eolla} for each \gls{thgr} $i$ as follows
\begin{equation}
    \eta^*_i = \arg \max_{\eta_i} \left(R_{\eta_i}| 1- P^{succ}_{\gls{thgr}_i}(\gamma_{eff}^{\gls{thgr}_i},\eta_i) \leq \text{\acrshort{tbler}\textsuperscript{T}}\right)
\end{equation}
where $P^{succ}_{\gls{thgr}_i}(\gamma_{eff}^{\gls{thgr}_i},\eta_i)$ is the probability of success of \gls{tb} decoding for \gls{thgr} $i$ after cooperation for \gls{mcs} index $\eta$.
This algorithm ensures that \gls{ptm} multicast transmission to a \gls{thgr} selects an \gls{mcs} that is at least as high as that used for a unicast transmission to \acrshort{ue-x} and typically higher when cooperation gains are present.
In the absence of cooperation gains, the selected \gls{mcs} matches that of the unicast case.

The proposed method extends the joint \gls{olla} algorithm of \cite{bib2-our3}, which supports only \gls{sc} cooperation.
Our enhanced version works with both \gls{scs} and \gls{sscs}, leveraging the additional second \gls{harq} feedback in \gls{sscs} and incorporating soft combining gain into \gls{la}.
\begin{algorithm}

\small
\caption{\small Enhanced JOINT \gls{olla}}\label{algo:olla1}
\begin{algorithmic}
\STATE $\{HF_{X1}, HF_{T1}\} \rightarrow $ 1\textsuperscript{st} \gls{harq} feedback from \{\acrshort{ue-x}, \acrshort{ue-t}\} \\
\STATE $HF_{X2} \rightarrow $ 2\textsuperscript{nd} \gls{harq} feedback from \acrshort{ue-x}  \\
\STATE $HF_{J} \rightarrow $ Joint \gls{harq} feedback input to \gls{olla} \\
\STATE $CS \rightarrow$ Cooperation Scheme (\gls{sc}, \gls{sscs}) 
\IF { $CS = $ \gls{sc} or \gls{ack} is received in 1\textsuperscript{st} \gls{harq} feedback}
\STATE $HF_J = HF_{X1} \vee HF_{T1}$ \\
\STATE where $\vee$ represents logical OR
\ELSE
\STATE $HF_J = HF_{X2}$
\ENDIF 
\end{algorithmic}
\end{algorithm}

The operation of joint \gls{harq} and enhanced joint \gls{olla} for different decoding scenarios of \gls{pdcch} and \gls{pdsch} (assuming Rank-1 transmission for each \gls{ptm} transmission to a \gls{thgr}) is summarized in Table \ref{tab:jointHO}.
It can be seen that joint \gls{harq} and enhanced joint \gls{olla} effectively utilize cooperation based reliability gains to increase spectral efficiency of the network by choosing higher \gls{mcs} values and avoiding redundant \gls{harq} retransmissions.
For \gls{scs}, \gls{harq} retransmissions are avoided in scenarios 2 and 3, with \gls{olla} step down ($\Delta_{down}$) is applied in both cases.
For \gls{sscs}, \gls{harq} retransmissions are additionally avoided in scenario 5 when \gls{ack} is received after soft combining, with \gls{olla} step down ($\Delta_{down}$) is applied in scenarios 2, 3 and 5.
\begin{table*}
\caption{\gls{pdcch} and \gls{pdsch} Decoding Status (Successfully Decoded: D and Fail to Decode: N/D) and Joint \gls{harq} \& \gls{olla}  for the Tethering Group with \gls{scs} and \gls{sscs}}
\begin{center}
\resizebox{2\columnwidth}{!}{%
\renewcommand{\arraystretch}{1.2}
\begin{tabular}{|c|c|c|c|c|c|c|c|c|c|c|c|c|}
\hline
 \textbf{Scenario} & \multicolumn{3}{|c|}{\textbf{\acrshort{ue-t}}}& \multicolumn{4}{|c|}{\textbf{\acrshort{ue-x}}} & \multicolumn{2}{|c|}{\textbf{\gls{scs}}} & \multicolumn{2}{|c|}{\textbf{\gls{sscs}}} \\
\hline
& \multirow{2}{*}{\textbf{\gls{pdcch}}}& \multirow{2}{*}{\textbf{\gls{pdsch}}}& \gls{harq} & \multirow{2}{*}{\textbf{\gls{pdcch}}}& \multirow{2}{*}{\textbf{\gls{pdsch}}} & \multicolumn{2}{|c|}{\gls{harq}} & \textbf{Joint \gls{harq}} & \textbf{Enhanced} & \textbf{Joint \gls{harq}} & \textbf{Enhanced}\\
& &  & Feedback & &  & \multicolumn{2}{|c|}{Feedback} & \gls{gNodeB} & \textbf{Joint \gls{olla}} & \gls{gNodeB} & \textbf{Joint \gls{olla}}\\
\cline{7-8}
& &  & & & & Initial & SoftC   & Retransmission &  & Retransmission & \\
\hline

1 & D & D & \gls{ack}  & D & D &\gls{ack} &- & \multirow{4}{*}{No} & \multirow{4}{*}{$\Delta_{down}$} & \multirow{5}{*}{No} & \multirow{5}{*}{$\Delta_{down}$}\\
\cline{1-8}
2 & D & D &\gls{ack} & D & N/D &\gls{nack} &-&  &  &  & \\
\cline{1-8}
3 & D & D &\gls{ack} & N/D & N/D &- &- &  &  &  & \\
\cline{1-8}
4 & D & N/D &\gls{nack} & D & D &\gls{ack} & -&  &  &  &  \\
\cline{1-8}
\cline{9-10}
\multirow{2}{*}{5} & \multirow{2}{*}{D} & \multirow{2}{*}{N/D} & \multirow{2}{*}{\gls{nack}} & \multirow{2}{*}{D} & \multirow{2}{*}{N/D} &\multirow{2}{*}{\gls{nack}} & \gls{ack} & \multirow{3}{*}{Yes} & \multirow{3}{*}{$\Delta_{up}$} &  &  \\
\cline{8-8}
\cline{11-12}
 &  &  &  &  &  & & \gls{nack} &  &  & \multirow{2}{*}{Yes} & \multirow{2}{*}{$\Delta_{up}$} \\
\cline{1-8}

6 & D & N/D &\gls{nack} & N/D & N/D &- &-&  &  &  & \\

\hline
7 & N/D & N/D &- & D & D &\gls{ack} & - & No & $\Delta_{down}$ & No & $\Delta_{down}$\\
\hline
8 & N/D & N/D &- & D & N/D &\gls{nack} & - &\multirow{2}{*}{Yes} & \multirow{2}{*}{$\Delta_{up}$} & \multirow{2}{*}{Yes} & \multirow{2}{*}{$\Delta_{up}$} \\
\cline{1-8}
9 & N/D & N/D &- & N/D & N/D &- &-&  &  &  &  \\
\hline
\end{tabular}
\label{tab:jointHO}
}
\end{center}
\end{table*}


\section{\gls{cb} based Cooperation}
\label{cbcoop}
We extend our investigation of \gls{thgr} cooperation by considering cooperation at the \gls{cb}s level instead of the full \gls{tb}. 
In this scenario, \acrshort{ue-t} forwards only the successfully decoded \gls{cb}s while applying the \gls{df} algorithm in case of \gls{scs}, along with associated signaling information indicating the indices of these \gls{cb}s.
The indexing overhead for \gls{cb}s is minimal, as discussed in \cite{bib2-our2}.
Meanwhile, \acrshort{ue-x} applies \gls{sc} at the \gls{cb} level if \gls{scs} is configured.

In \gls{sscs} configuration, \acrshort{ue-t} forwards both the decoded \gls{cb}s and the soft bits of the \gls{cb}s it fails to decode.
It this configuration, it also forwards the two sets of index information, one for the decoded \gls{cb}s and another for the \gls{cb}s whose soft bits are forwarded.
And \acrshort{ue-x} decides whether to apply \gls{sc} or soft combining based on its own decoding status of each \gls{cb}. 
If a \gls{cb} is already decoded by \acrshort{ue-x}, any corresponding data received from \acrshort{ue-t} is discarded.
If the \gls{cb} is not decoded during the initial decoding attempt, \acrshort{ue-x} applies \gls{sc} if it receives the decoded \gls{cb} from \acrshort{ue-t} or performs soft combining if it receives the corresponding soft bits.

For this type of cooperation, we also incorporate \gls{cbg} based \gls{harq} retransmissions from the \gls{gNodeB}.
In this setup, \gls{harq} retransmissions are \gls{cbg} based.
However, \gls{olla} algorithm remains based on \gls{harq} feedback at the \gls{tb} level.
To encourage higher \gls{mcs} selection with this setup, we configure the target \gls{bler} to 30\% (rather than the typical 10\%) as highlighted in \cite{bib4-4}. 
Despite this higher target, the actual \gls{bler} remains below the target value. 
The functionality of joint \gls{olla} is unaffected, as it continues to operate on \gls{tb}-level \gls{harq} feedback.
For joint \gls{harq} per \gls{cbg} the logic remains the same, except that both \gls{ue}s would now provide \gls{harq} feedback per \gls{cbg} instead of per \gls{tb}.

\subsection{Limited \gls{cb} based \gls{sscs}}
We consider two limited \gls{cb}-based \gls{sscs} schemes, previously proposed in our prior work in \cite{bib2-our2}.
These schemes offer a trade-off between performance gains and the volume of data exchanged over the tethering link.
Specifically, reducing the percentage of \gls{cb}s shared for soft combining leads to reduced performance compared to sharing 100\% of the \gls{cb}s.
In both schemes, if both \gls{ue}s per \gls{thgr} failed to decode the \gls{tb}, soft combining is applied only on a subset of \gls{cb}s corresponding to that \gls{tb}. 
This subset is selected based on relatively lower \gls{sinr} values among the \gls{cb}s. 
The soft bits corresponding to these \gls{cb}s can either be requested by \acrshort{ue-x} or proactively forwarded by \acrshort{ue-t}, depending upon the scheme.
Any \gls{cb}s that remain undecoded after soft combining are retransmitted by the \gls{gNodeB} using \gls{cbg} based \gls{harq}.
\begin{itemize}
    \item \gls{cb}s-\acrshort{ue-x}: In this scheme, \acrshort{ue-x} identifies a limited percentage of \gls{cb}s (based on their low \gls{sinr} values) from the \gls{tb} that it failed to decode and requests the corresponding soft bits from \acrshort{ue-t} over the tethering link. \acrshort{ue-x} sends the indexing information as specified in \cite{bib2-our2} and \acrshort{ue-t} responds by sending only the requested soft bits.
    \item \gls{cb}s-\acrshort{ue-t}: In this scheme, \acrshort{ue-t} proactively sends soft bits for a limited percentage of \gls{cb}s from the \gls{tb}, that it failed to decode, along with their indexing information \cite{bib2-our2}.\acrshort{ue-x} then applies soft combining on these limited \gls{cb}s if it was also unable to decode that \gls{cb} initially. 
\end{itemize}

\section{Methodology for Performance Evaluation}
\label{evam}
We employ a dynamic \gls{sls}, aligned with the methodology described in the \gls{sls} tutorial \cite{bib3-8}, to evaluate the proposed \gls{xr} \gls{thgr} framework.
The simulation setup adheres to the system model detailed in Section \ref{systemmodel}, with parameters summarized in Table \ref{tabpara}. 
This simulator was also utilized in our previous study \cite{bib2-our3}.
Calibration of the \gls{sls} is performed using the \gls{3gpp} \gls{nr} Release-18 \gls{xr} evaluation guidelines as specified in \cite{bib2}.

The simulator simulates a comprehensive protocol stack, including application, transport, internet protocol, packet data convergence protocol, radio link control, medium access control, and physical layers.
It supports key \gls{rrm} capabilities such as \gls{gNodeB} scheduler, retransmission from the \gls{gNodeB} based on \gls{harq} feedback, \gls{csi} measurement and periodic reporting, and \gls{la} (both inner and outer loop), etc.
Consistent with the \gls{5g} \gls{nr} \gls{tdd} structure, the simulator implements a slot-based transmission model using 14 \gls{ofdm} symbols per slot with a normal cyclic prefix \cite{bib16}.
We follow the DDDSU \gls{tdd} frame structure in line with \gls{3gpp} \gls{nr} \gls{xr} evaluation criteria.
It represents three \gls{dl} slots, a special slot comprising of 10 \gls{dl} symbols, two symbols as guard period and two symbols for \gls{ul} and lastly one complete \gls{ul} slot.

For each \gls{thgr}, the \gls{sinr} is calculated for both \gls{ue}s once \gls{dl} \gls{ptm} multicast transmission is scheduled depending upon the \gls{xr} traffic for the \acrshort{ue-x} of that \gls{thgr}.
The \gls{sinr}  calculations accounts for \gls{svd} precoder at the transmitter, \gls{mimo} configurations, channel effects, and the \gls{mmseirc} receiver.
The \gls{mmib} is then derived for both \gls{ue}s of the \gls{thgr} utilizing their \gls{sinr}\cite{bib2-10}.
The block error probability, which is obtained from the received \gls{sinr}, \gls{mcs} and \gls{mmib} via a precomputed look-up table produced from thorough link-level simulations, is used to determine the decoding success or failure for each \gls{ue} of that \gls{thgr}.
%

The \gls{gNodeB} schedules transmissions in following order
\begin{enumerate}
    \item \gls{harq} retransmissions (if any for \gls{xr})
    \item \gls{xr} traffic
    \item \gls{harq} retransmissions (if any for \gls{embb})
    \item \gls{embb} traffic
\end{enumerate}
The \gls{gNodeB} schedules transmissions 3 and 4 only in scenarios where both traffic types coexists.

\subsection{Deployment Scenario}

\begin{figure}
\centerline{\includegraphics[width=9cm]{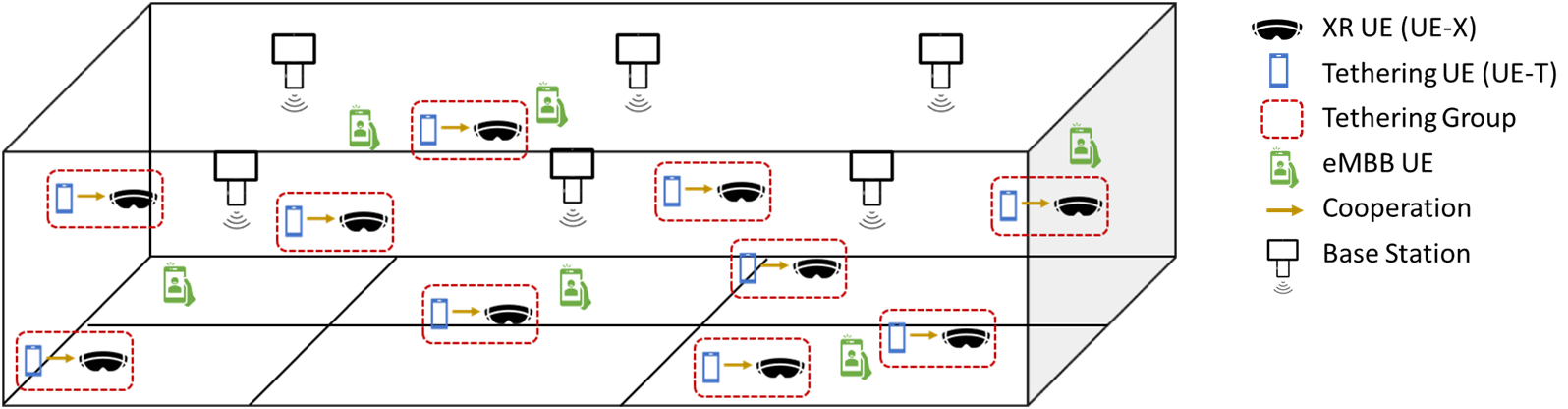}}
\caption{Indoor hotspot scenario with \gls{xr} \gls{thgr}s.}
\label{fig:InH}
\end{figure}

\begin{figure}
\centerline{\includegraphics[width=9cm]{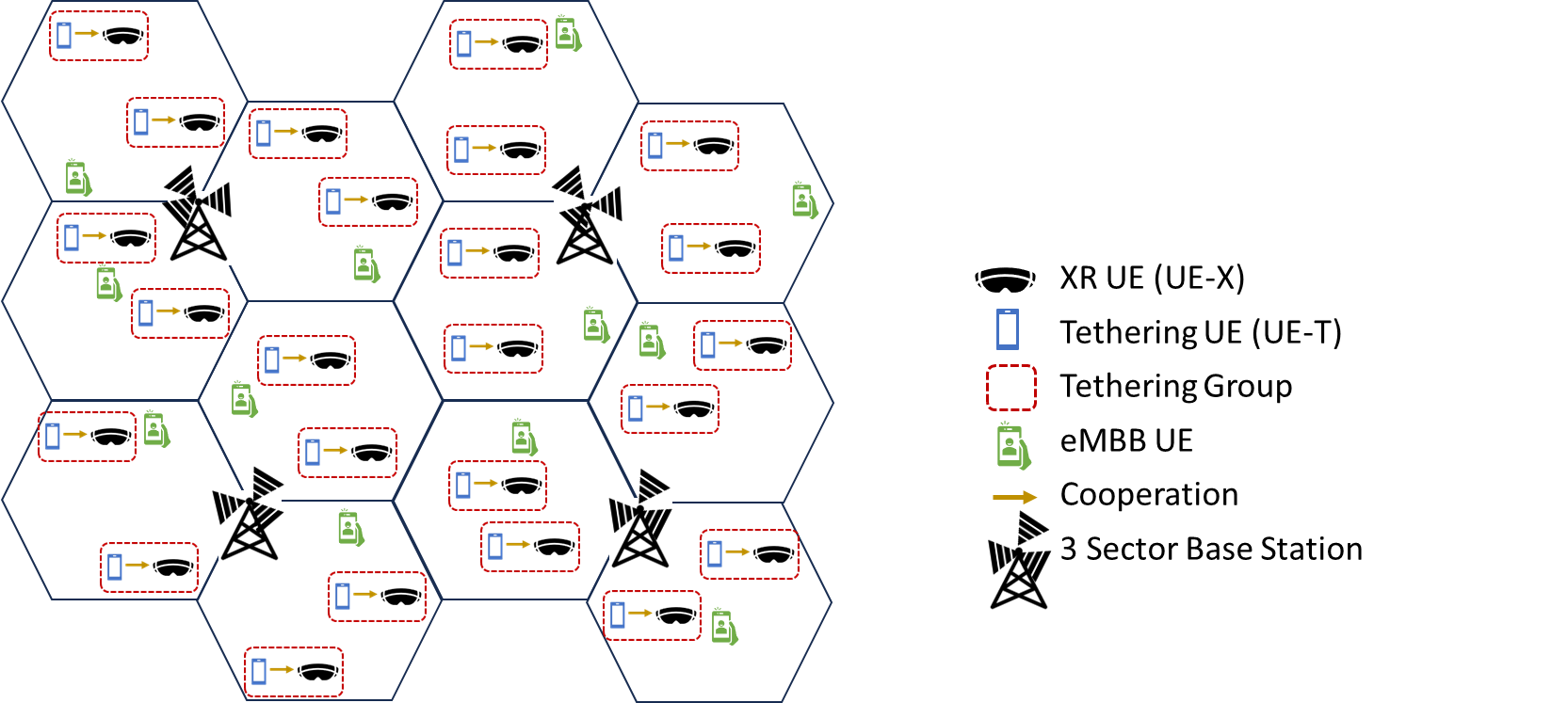}}
\caption{Dense Urban scenario with \gls{xr} \gls{thgr}s.}
\label{fig:DU}
\end{figure}

This study evaluates the system performance under the following two standardized \gls{3gpp} defined deployment scenarios:

\begin{enumerate}
    \item \gls{inh}: This scenario models an indoor environment measuring 120 $\times$ 50 meters, with all users located indoors as shown in Fig. \ref{fig:InH}. 
    The deployment includes 12 single-cell base stations, with 6 positioned in each of two parallel rows, separated by 20 meters of inter-site distance.
    The base stations are ceiling-mounted and operate at a lower transmit power relative to outdoor deployments.
    \item \gls{du}: This scenario represents an outdoor dense urban environment spanning 528 $\times$ 60 meters, incorporating three-sector 7 base stations with a total of 21 cells as illustrated in Fig. \ref{fig:DU}. Buildings are randomly placed and it is assumed that 80\% of the users are located indoors.
     
\end{enumerate}
The channel characteristics for both deployment scenarios follow the \gls{3gpp}-defined models in \cite{bib13}.

\subsection{Traffic Model}
The specification in \cite{bib2} is followed for generating the \gls{xr} traffic.
\gls{xr} video frames of varying sizes are generated using truncated Gaussian distribution at a specified rate of 60 frames per second. 
The arrival of these frames at the \gls{gNodeB} are modeled through a distinct truncated Gaussian distribution.
$\mathcal{TN}(\mu,\sigma,a,b)$ is the notation for a truncated Gaussian distribution in Table \ref{tabpara}, where $\mu$ and $\sigma$ stand for the mean and standard deviation and $a$ and $b$ for the minimum and maximum bounds.
\gls{3gpp} in \cite{bib2} defined the \gls{kpi}, \gls{xr} capacity, for measuring the capacity of the network with \gls{xr} users, i.e., the maximum number of \gls{xr} \gls{ue}s that may be served per cell when at least 90\% of them are happy.
An \gls{xr} user is considered happy if it successfully receive 99\% of the frames (packets) within the \gls{pdb}.

For coexistence scenarios, we consider one \gls{embb} user with full buffer traffic per cell in addition to the\gls{xr} users.
This indicates that there is unlimited data intended for the \gls{embb} \gls{ue} at the \gls{gNodeB} waiting for scheduling and subsequent transmission.

For baseline simulations, only \gls{xr} \gls{ue}s are dropped in the network and the \gls{gNodeB} transmits \gls{xr} data via \gls{ptp} (unicast) to each \gls{xr} \gls{ue}.
These \gls{ue}s have only single link to the network and are referred to as legacy \gls{xr} \gls{ue}s.  
Additionally, in case of coexistence scenario baseline simulations also include one \gls{embb} \gls{ue} per cell.
\begin{table}
\caption{Simulation Parameters}
\begin{center}

\begin{tabular}{c|c|c}
\hline
\textbf{Parameter} & \multicolumn{2}{|c}{\textbf{Setting}} \\
Deployment & \gls{inh} & \gls{du}\\
\hline
Area & 120m $\times$ 50m & 528m $\times$ 460m \\
\hline
Layout & 12 cells & 21 cells\\
\hline
Inter-site Distance & 20m & 200m\\
\hline
\gls{gNodeB} height  & 3m & 25m\\ 
\hline
\gls{gNodeB} power & 31 dBm & 51 dBm\\
\hline
Indoor \gls{ue} Probability & 1 & 0.8 \\
\hline
Number of building floors & 1 & 6 \\
\hline
\gls{ue} distribution among floors & & uniform(1,6) \\
\hline
Simulation Time & \multicolumn{2}{c}{9 seconds}\\
\hline
Simulation runs & \multicolumn{2}{c}{10} \\
\hline
TTI length & \multicolumn{2}{c}{14 \gls{ofdm} symbols}\\
\hline
\multirow{1}{*}{\gls{tdd} Frame Structure} & \multicolumn{2}{c}{DDDSU}\\
\hline
Bandwidth & \multicolumn{2}{c}{100 MHz} \\
\hline
Carrier Frequency & \multicolumn{2}{c}{4 GHz} \\
\hline
sub-carrier spacing & \multicolumn{2}{c}{30 kHz} \\
\hline
Modulation & \multicolumn{2}{c}{QPSK to 256QAM}  \\
\hline
\gls{xr} frame rate & \multicolumn{2}{c}{60 fps} \\
\hline
\gls{xr} random jitter & \multicolumn{2}{c}{$\mathcal{TN}(0,2,-4,4)$ ms}\\
\hline
\gls{xr} frame size (45 Mbps) & \multicolumn{2}{c}{$\mathcal{TN}(93,10,46,141)$ kB}\\
\hline

\multirow{2}{*}{\gls{gNodeB} antenna} & \multicolumn{2}{c}{1 panel with 32 elements}\\
& \multicolumn{2}{c}{(4 $\times$ 4 and 2 polarization)} \\
\hline
\gls{gNodeB} \acrshort{tx} processing delay & \multicolumn{2}{c}{2.75 \gls{ofdm} symbols} \\
\hline
\gls{ue} height &  \multicolumn{2}{c}{1.5m} \\
\hline
\multirow{2}{*}{\gls{ue} antenna} & \multicolumn{2}{c}{4 omni-directional antennas} \\
 & \multicolumn{2}{c}{(2 $\times$ 1 and 2 polarization)}\\
\hline

\gls{ue} receiver & \multicolumn{2}{c}{\gls{mmseirc}} \\
\hline
\gls{ue} speed & \multicolumn{2}{c}{3 km/h} \\
\hline
\gls{ue} \acrshort{rx} processing delay & \multicolumn{2}{c}{6 \gls{ofdm} symbols}\\
\hline
Scheduler & \multicolumn{2}{c}{Proportional Fair} \\
\hline
\gls{cqi} Measurement & \multicolumn{2}{c}{Periodic every 2 ms}\\
\hline
\gls{cqi} Reporting Delay & \multicolumn{2}{c}{2 ms} \\
\hline
Rank & \multicolumn{2}{c}{1} \\
\hline
Channel Estimation & \multicolumn{2}{c}{Ideal} \\
\hline
\gls{harq} combining method & \multicolumn{2}{c}{Chase soft combining} \\
\hline
Target \gls{bler} & \multicolumn{2}{c}{10\%}\\
\hline
Intra \gls{thgr} distance ($x$) & \multicolumn{2}{c}{1m} \\
\hline
\end{tabular}

\label{tabpara}
\end{center}
\end{table}

Each simulation runs for 9 seconds, enabling the evaluation of \gls{xr} users satisfaction by verifying that 99\% of packets are received within the \gls{pdb} with a 99\% confidence level and an error margin of 1.1\%. 
The confidence level is calculated using the similar analogy as used in \cite{bib3-6}, i.e., 9 seconds simulation generates 540 packets per \gls{xr} \gls{ue}, which under the assumption of uncorrelated packets provides the required statistical confidence for the happiness metric. 
Simulations are repeated 10 times with different seeds to capture diverse user locations and channel conditions.
For instance, with 7 \gls{xr} \gls{ue}s (or \gls{thgr}) per cell, the system collect statistics from 840 \gls{xr} \gls{ue}s in \gls{inh} scenario (7 $\times$ 10 drops $\times$ 12 cells)  and 1470 \gls{xr} \gls{ue}s in \gls{du} scenario (7 $\times$ 10 drops $\times$ 21 cells). 
This approach facilitates estimation of \gls{xr} capacity, with 7 \gls{xr} \gls{ue}s (or \gls{thgr}) per cell, achieving 99\% confidence level for an error margin of 2.6\% for \gls{inh} and for an error margin of 2\% for \gls{du} scenario.
The simulations are further repeated for varying numbers of \gls{xr} \gls{ue}s (or \gls{thgr}s) per cell, using uniformly distributed user locations and uniform load per cell.

\section{Simulation Results and Discussion}
\label{simrst}
\subsection{\gls{tb} based Cooperation}
\subsubsection{\gls{xr} only}

\begin{figure}
     \centering
     \begin{subfigure}[b]{0.5\textwidth}
         \centering
         \includegraphics[width=8.5cm]{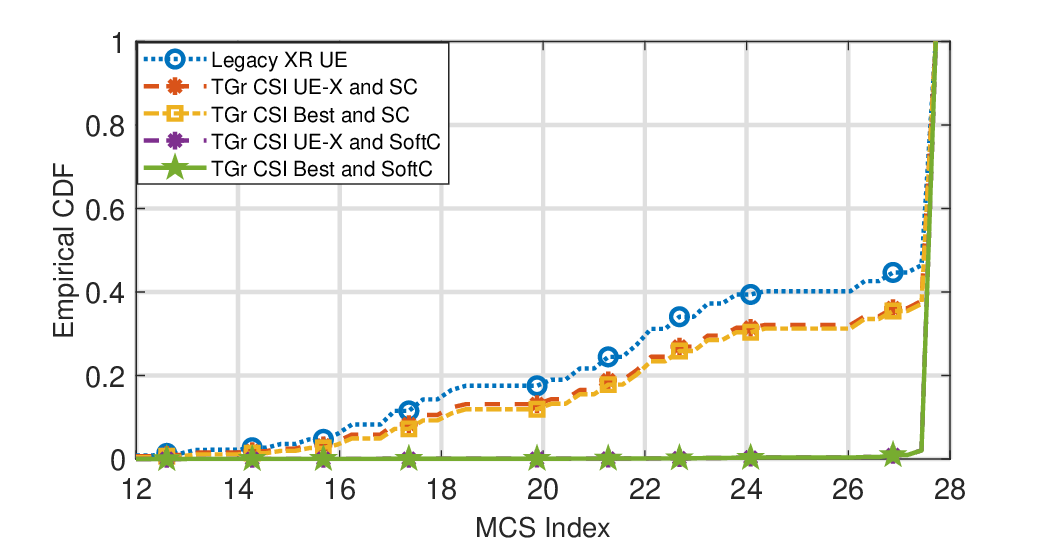}
         \caption{\gls{inh}}
         \label{fig:mcsinh}
     \end{subfigure}
     \hfill
     \begin{subfigure}[b]{0.5\textwidth}
         \centering
         \includegraphics[width=8.5cm]{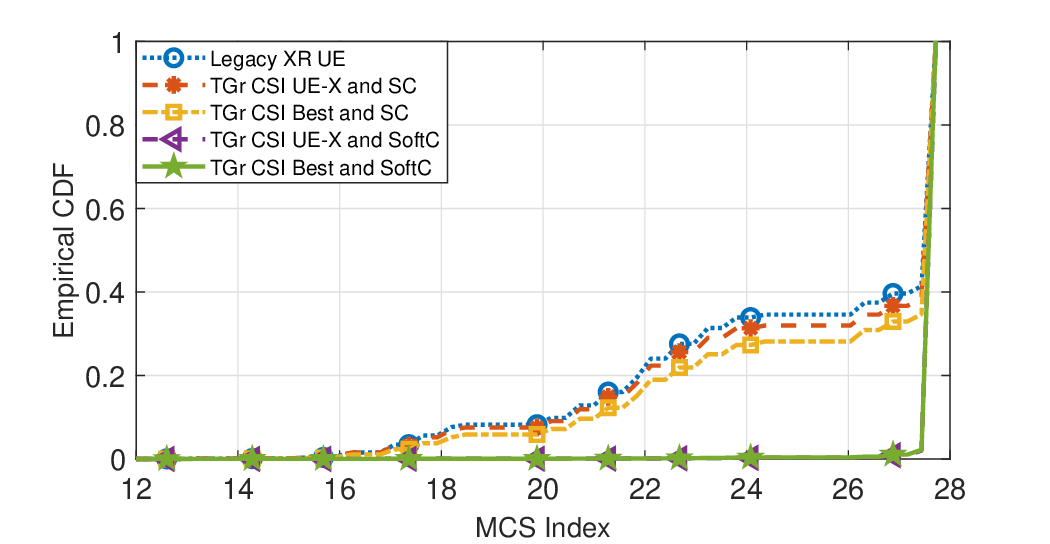}
         \caption{\gls{du}}
         \label{fig:mcsdu}
     \end{subfigure}
     \hfill
        \caption{MCS Index with 7 XR UEs per cell. (a) \gls{inh} (b) \gls{du}.}
        \label{fig:MCS}
\end{figure}

Initial simulations were conducted with only \gls{xr} \gls{ue}s / \gls{thgr}s and the results are summarized in this subsection.

Figure \ref{fig:MCS} presents the \gls{mcs} index selection as an \gls{ecdf}.
The results are presented for \gls{xr} \gls{thgr}s using both \gls{sc} and \gls{sscs}, across two variations of \gls{csi} reporting per group (as discussed in Section \ref{systemmodel}).
For reference, the performance of legacy \gls{xr} \gls{ue}s is also shown.
These results corresponds to scenarios with 7 \gls{xr} users (legacy or \gls{thgr}) per cell. 
The \gls{xr} \gls{thgr}s exhibit significantly improved \gls{mcs} selection due to the proposed enhanced joint \gls{olla} mechanism.
Among the cooperation schemes, \gls{sscs} consistently delivers superior performance compared to \gls{sc}.
For instance,  under the given simulation conditions, \gls{xr} \gls{thgr} with \gls{sscs} (for both \gls{csi} variations) select the highest available \gls{mcs} index in approximately 98\% of simulation time in both \gls{du} and \gls{inh} scenarios. 
In contrast, \gls{xr} \gls{thgr}s with \gls{sc} and \gls{csi} best select the highest \gls{mcs} 65\% of the time in \gls{du} and 63\% in \gls{inh}.
The legacy \gls{xr} \gls{ue}s achieve this only 58\% and 54\% of the time in \gls{du} and \gls{inh}, respectively.
The impact of the two \gls{csi} reporting variations on \gls{sscs} performance with proposed enahcned joint \gls{olla} algorithm is minimal across both deployment scenarios.
However, the performance difference due to \gls{csi} reporting variations is more visible for \gls{scs} in \gls{du} scenario compared to \gls{inh}. 
This is attributed to greater \gls{sinr} variability between the closely located \acrshort{ue-x} and \acrshort{ue-t} within a \gls{thgr} in \gls{du} compared to \gls{inh}.
As shown in Table \ref{tab:ps}, the \gls{du} scenario exhibits a higher proportion of \gls{thgr}s with \gls{ue}s experiencing distinct propagation conditions (\gls{nlos} / \gls{los}), contributing to greater \gls{sinr} imbalance compared to \gls{inh}.
Nevertheless, under \gls{sscs}, performance remains consistent across \gls{csi} reporting variation.

\begin{table}
\caption{Propagation Status (PS) [\%] of \gls{ue}s within \gls{thgr} with 7 \gls{xr} \gls{thgr}s/cell}
\begin{center}
\resizebox{0.9\columnwidth}{!}{%

\begin{tabular}{|c|c|c|c|c|}
\hline
\multirow{2}{*}{\textbf{Scenario}}&\multicolumn{2}{|c|}{\textbf{\gls{thgr} PS}}& \multirow{2}{*}{\textbf{\gls{du}}} & \multirow{2}{*}{\textbf{\gls{inh}}}\\
\cline{2-3} 
 & \textbf{\acrshort{ue-x}}& \textbf{\acrshort{ue-t}} &  &  \\
\hline
1 & \gls{los} & \gls{los} & 48.3 & 98.8\\
\hline
2 & \gls{los} & \gls{nlos} & 3.7 & 0\\
\hline
3 & \gls{nlos} & \gls{los} & 6.8 & 1.2\\
\hline
4 & \gls{nlos} & \gls{nlos} & 41.2 & 0\\
\hline
\end{tabular}
\label{tab:ps}
}
\end{center}
\end{table}

\begin{figure}
     \centering
     \begin{subfigure}[b]{0.5\textwidth}
         \centering
         \includegraphics[width=8.5cm]{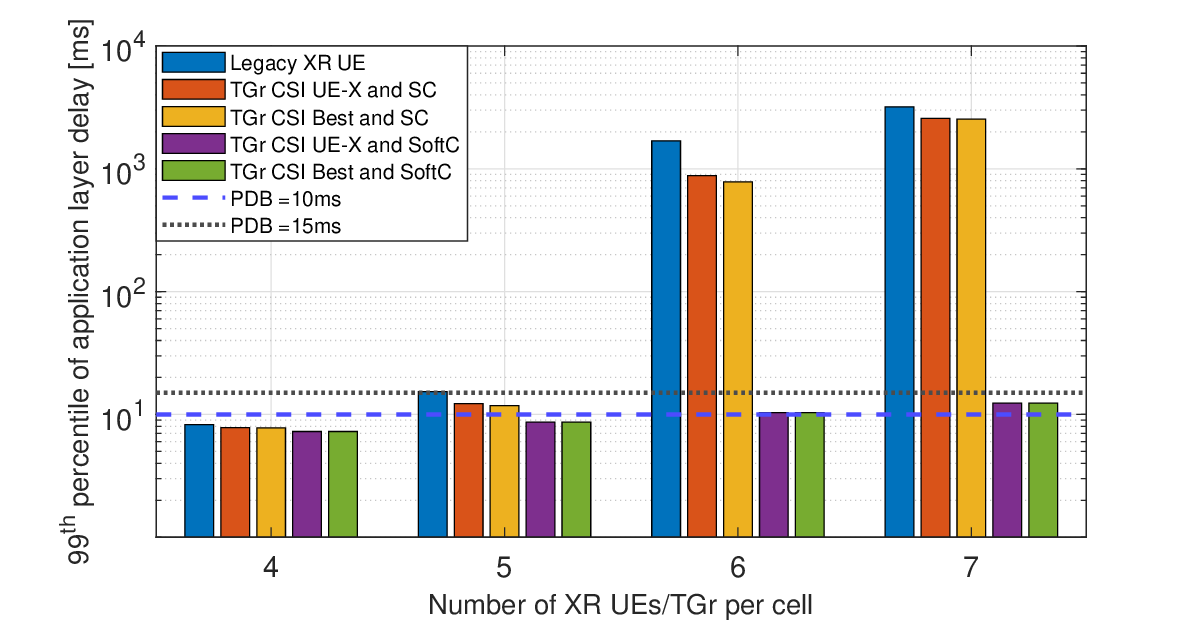}
         \caption{\gls{inh}}
         \label{fig:delayinh}
     \end{subfigure}
     \hfill
     \begin{subfigure}[b]{0.5\textwidth}
         \centering
         \includegraphics[width=8.5cm]{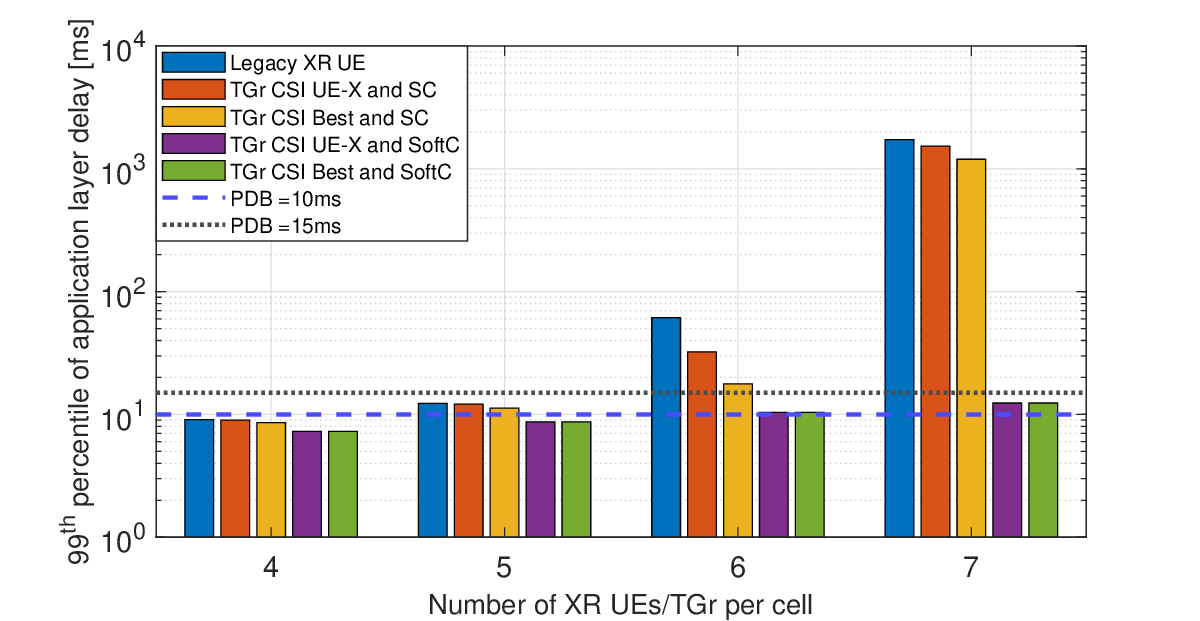}
         \caption{\gls{du}}
         \label{fig:delaydu}
     \end{subfigure}
     \hfill
        \caption{99\textsuperscript{th} percentile of \gls{ecdf} of application layer delay. (a) \gls{inh} (b) \gls{du}.}
        \label{fig:delay}
\end{figure}

Figure \ref{fig:delay} illustrates the impact of \gls{thgr} cooperation (\gls{sc} and \gls{sscs}) with \gls{ptm} transmission on application layer delay for varying numbers of \gls{xr} \gls{ue}s or \gls{thgr}s per cell.
The figure shows the 99\textsuperscript{th} percentile of the \gls{ecdf} of application layer delay.
At lower user densities, both legacy \gls{xr} \gls{ue}s and \gls{xr} \gls{thgr}s experience similar delays.
This is due to favorable \gls{sinr} conditions and consequently, operation at maximum available \gls{mcs} by both legacy \gls{xr} \gls{ue}s and \gls{xr} \gls{thgr}.
\gls{thgr}s offer no additional benefit because the \gls{mcs} ceiling limits further gains.
As user densities increases, network interference rises, degrading \gls{sinr} and leading to lower \gls{mcs} selection for legacy \gls{xr} \gls{ue}s.
In contrast, \gls{thgr}s employing \gls{sc} or \gls{sscs} can sustain higher \gls{mcs} values, thereby significantly reducing application layer delay, where \gls{sscs} offers higher gains.
The interference impact is more severe in \gls{inh} than in \gls{du}. 
For example, at 6 and 7 \gls{xr} \gls{ue}s per cell,  legacy \gls{xr} \gls{ue}s in \gls{inh} suffers considerably higher delay than in \gls{du}.
However, \gls{xr} \gls{thgr} with \gls{sscs} maintain much lower delay is both deployments by operating at maximum \gls{mcs} 98\% of the time.
This confirms the robustness of \gls{sscs} under high interference scenarios (or low-\gls{sinr} conditions). 
\gls{thgr} with \gls{scs} offer better performance than legacy \gls{xr} \gls{ue}s but fall short of the gains achieved by \gls{sscs} in lower \gls{sinr} scenarios. 
The impact of \gls{csi} reporting variation on delay trends aligns with that observed in \gls{mcs} index selection.


\begin{figure}
     \centering
     \begin{subfigure}[b]{0.5\textwidth}
         \centering
         \includegraphics[width=8.5cm]{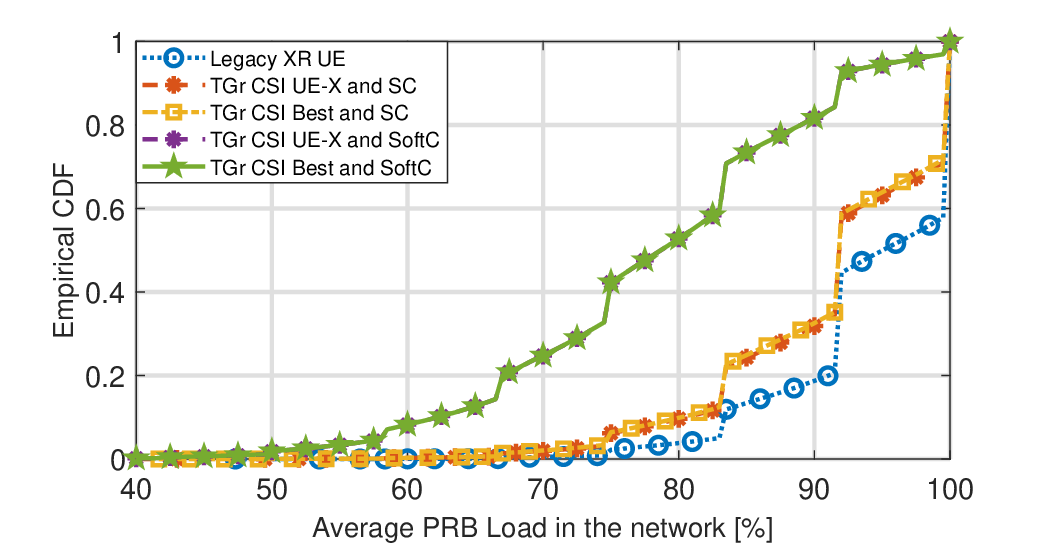}
         \caption{\gls{inh}}
         \label{fig:prbinh}
     \end{subfigure}
     \hfill
     \begin{subfigure}[b]{0.5\textwidth}
         \centering
         \includegraphics[width=8.5cm]{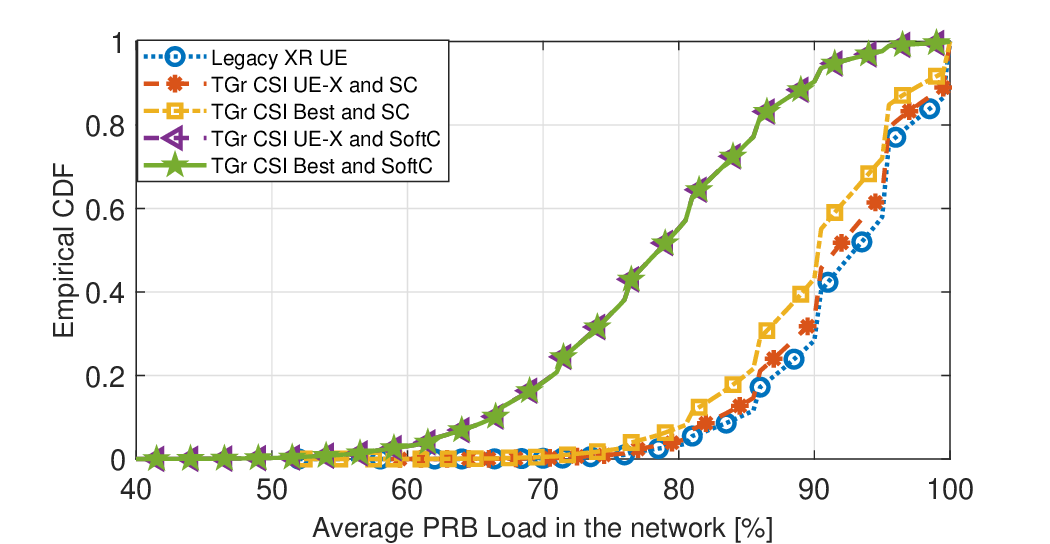}
         \caption{\gls{du}}
         \label{fig:prbdu}
     \end{subfigure}
     \hfill
        \caption{Average PRB Load in the network with 7 \gls{xr} \gls{ue}s per cell. (a) \gls{inh} (b) \gls{du}.}
        \label{fig:prb}
\end{figure}

Figure \ref{fig:prb} shows the average \gls{prb} utilization in the network when 7 \gls{xr} \gls{ue}s or \gls{thgr}s are deployed per cell in both \gls{du} and \gls{inh} scenarios.
\gls{ptm} multicast transmission with joint \gls{harq} and enhanced joint \gls{olla} to \gls{thgr}s reduces \gls{prb} usage compared to the unicast transmission to legacy \gls{xr} \gls{ue}s.
Specifically, \gls{thgr} with \gls{sscs} utilize 14-16\% fewer \gls{prb}s than legacy \gls{xr} \gls{ue}s at the 50\textsuperscript{th} percentile, while \gls{thgr} with \gls{scs} can save 1-3\% \gls{prb}s.
These gains stem from the ability of \gls{xr} \gls{thgr} to operate at higher \gls{mcs} levels and lower \gls{bler}s (or fewer \gls{harq} retransmissions) than legacy \gls{xr} \gls{ue}s for given \gls{sinr}. 
The extra \gls{prb}s can either support more \gls{xr} user(s) or be allocated to other traffics in coexistence scenarios.
\begin{figure}
     \centering
     \begin{subfigure}[b]{0.5\textwidth}
         \centering
         \includegraphics[width=8.5cm]{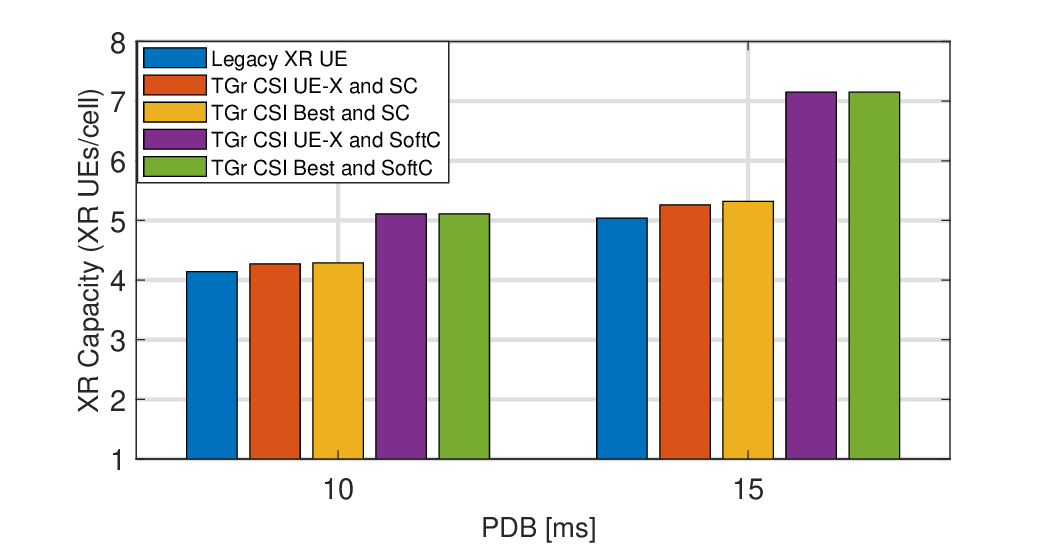}
         \caption{\gls{inh}}
         \label{fig:xrcinh}
     \end{subfigure}
     \hfill
     \begin{subfigure}[b]{0.5\textwidth}
         \centering
         \includegraphics[width=8.5cm]{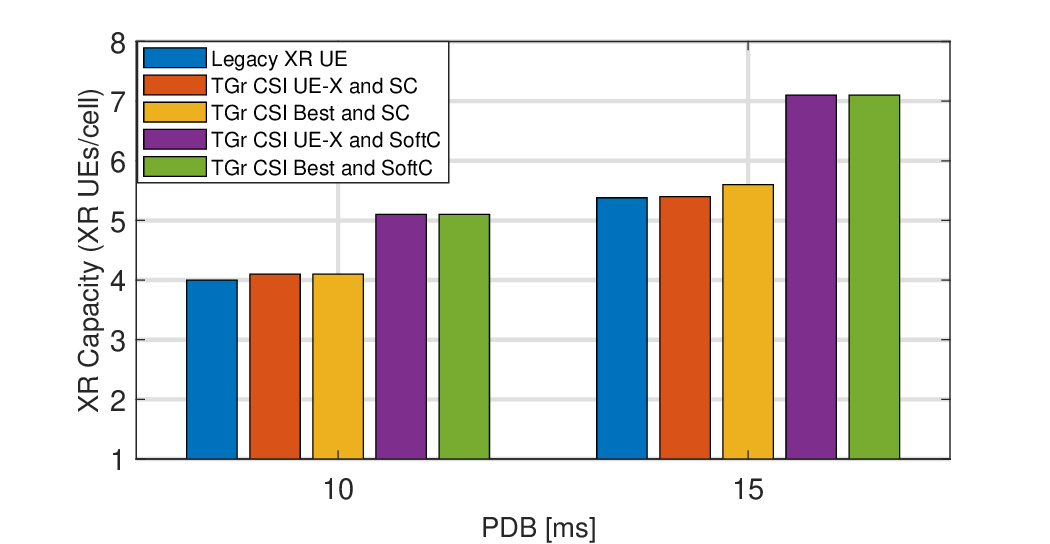}
         \caption{\gls{du}}
         \label{fig:xrcdu}
     \end{subfigure}
     \hfill
        \caption{XR Capacity (XR users only) vs PDB. (a) \gls{inh} (b) \gls{du}.}
        \label{fig:xrcapxronly}
\end{figure}

Figure \ref{fig:xrcapxronly} presents the \gls{xr} capacity for \gls{xr} \gls{thgr} with \gls{sc} and \gls{sscs} cooperation schemes alongside the baseline legacy \gls{xr} \gls{ue}s.
The \gls{xr} capacity is calculated with two \gls{pdb} thresholds, i.e., 10ms and 15ms.
The ability of \gls{xr} \gls{thgr} to operate at a higher \gls{mcs} for a given \gls{sinr} offers significant advantages over legacy \gls{xr} \gls{ue}s.
Most notably, a substantial reduction in \gls{prb} utilization and application layer delay.
These combined gains directly translate into a marked improvement in \gls{xr} capacity, highlighting the benefits of the \gls{thgr}-based cooperative approach under \gls{ptm} transmission with joint \gls{harq} and enhanced joint \gls{olla}.
\gls{xr} \gls{thgr} with \gls{sscs} outperforms both \gls{xr} \gls{thgr} with \gls{sc} and legacy \gls{xr} \gls{ue}.
For instant, \gls{thgr} with \gls{sscs} offer \gls{xr} capacity improvements of 23-27\% (\gls{pdb} = 10ms) and 32-42\% (\gls{pdb} = 15ms) over legacy \gls{xr} \gls{ue}s in both deployment scenarios.
In contrast, \gls{thgr} with \gls{scs} show modest capacity gains of 2.5-4\% under both \gls{pdb} targets.
The two variations of \gls{csi} per \gls{thgr} resulted in smaller performance difference with \gls{sscs} cooperation scheme compared to \gls{scs}.
Furthermore, the performance difference of \gls{xr} \gls{thgr} with \gls{scs} for two \gls{csi} variations is less pronounced in \gls{inh} scenario compared to \gls{du}.


\subsubsection{\gls{xr} \gls{ue}s plus one \gls{embb} \gls{ue}}
In the coexistence scenario, one \gls{embb} \gls{ue} with full buffer traffic per cell is added alongside the existing \gls{xr} \gls{ue}s / \gls{thgr}s per cell.
This setup enhances interference to \gls{xr} users, under the scheduling constraint that \gls{xr} traffic is prioritized, followed by \gls{embb} traffic.

\begin{figure}
     \centering
     \begin{subfigure}[b]{0.5\textwidth}
         \centering
         \includegraphics[width=8.5cm]{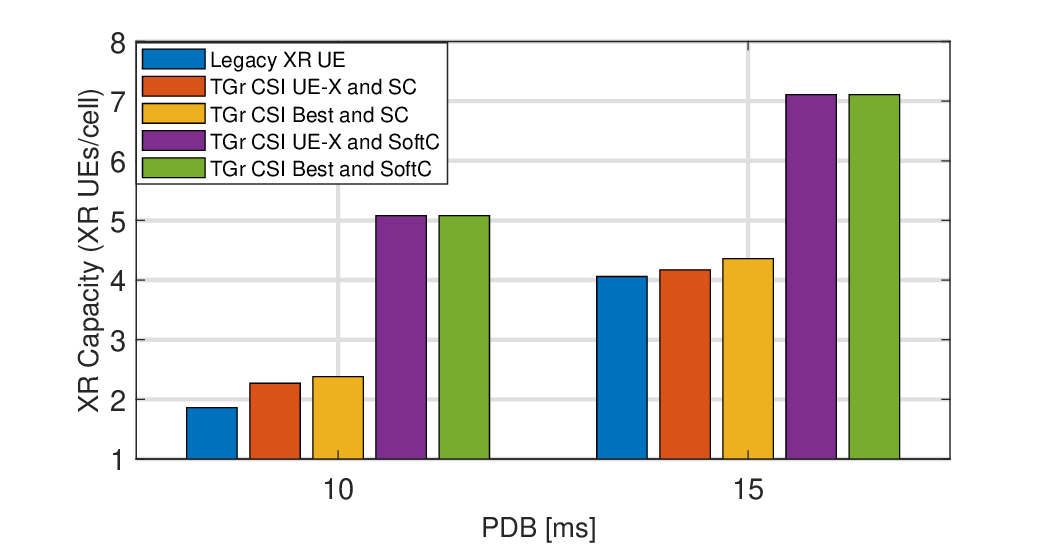}
         \caption{\gls{inh}}
         \label{fig:xrcinh}
     \end{subfigure}
     \hfill
     \begin{subfigure}[b]{0.5\textwidth}
         \centering
         \includegraphics[width=8.5cm]{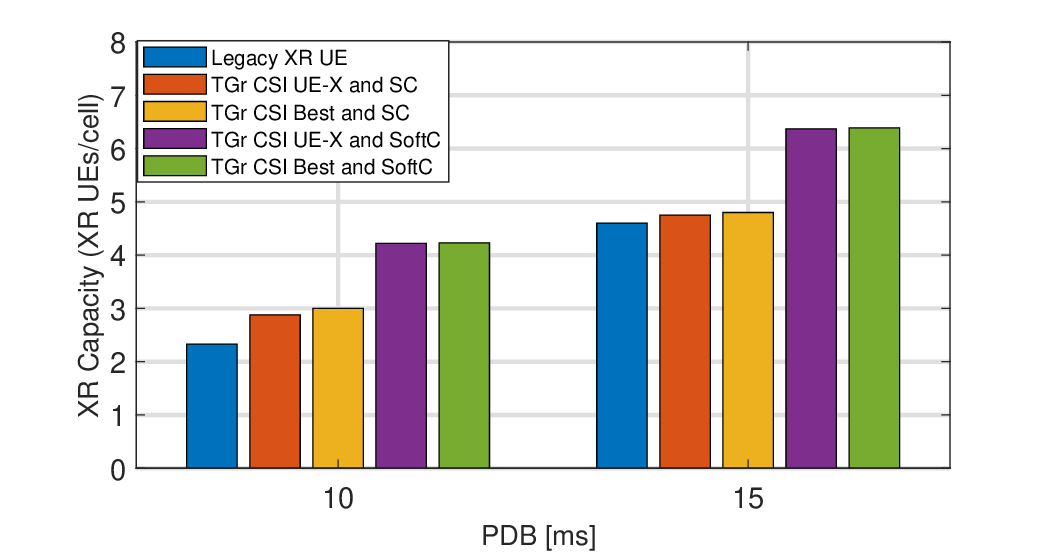}
         \caption{\gls{du}}
         \label{fig:xrcdu}
     \end{subfigure}
     \hfill
        \caption{XR Capacity (XR users plus 1eMBB user) vs PDB. (a) \gls{inh} (b) \gls{du}.}
        \label{fig:xrcapxrplusembb}
\end{figure}

Figure \ref{fig:xrcapxrplusembb} illustrates the \gls{xr} capacity of the network for two \gls{pdb} values (10ms and 15ms).
As expected, the \gls{xr} capacity of the network drops compared to \gls{xr}-only scenario (network with \gls{xr} \gls{ue}s / \gls{thgr}s).
The addition of a \gls{embb} \gls{ue} with full buffer traffic per cell increases the interference in the system, which ultimately reduces \gls{xr} capacity.
However, the relative performance gains of \gls{xr} \gls{thgr}s are higher than in the \gls{xr}-only case.
Specifically, in the \gls{inh} scenario, \gls{xr} \gls{thgr} achieve gains of 3-28\% with \gls{scs} and 75-173\% with \gls{sscs}, while in \gls{du} the gains are 3-29\% with \gls{scs} and 38-82\% with \gls{sscs}.
Notably, in \gls{inh}, the performance of \gls{xr} \gls{thgr} with \gls{sscs} remains largely unaffected by the added \gls{embb} interference, maintaining nearly the same performance as in the \gls{xr}-only case and outperforming legacy \gls{xr} \gls{ue}s by a greater margin.

\begin{figure}
     \centering
     \begin{subfigure}[b]{0.5\textwidth}
         \centering
         \includegraphics[width=8.5cm]{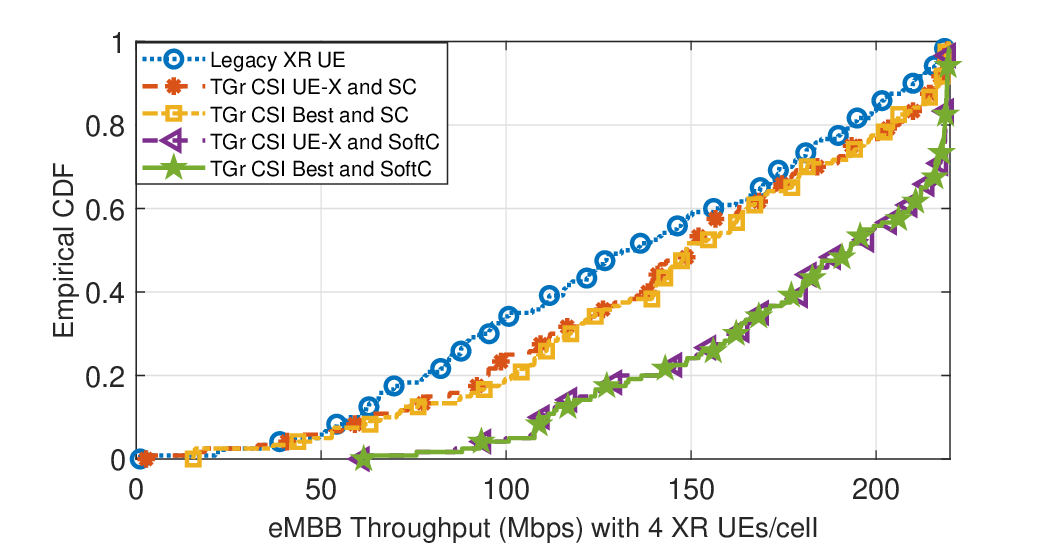}
         \caption{\gls{inh}}
         \label{fig:xrcinh}
     \end{subfigure}
     \hfill
     \begin{subfigure}[b]{0.5\textwidth}
         \centering
         \includegraphics[width=8.5cm]{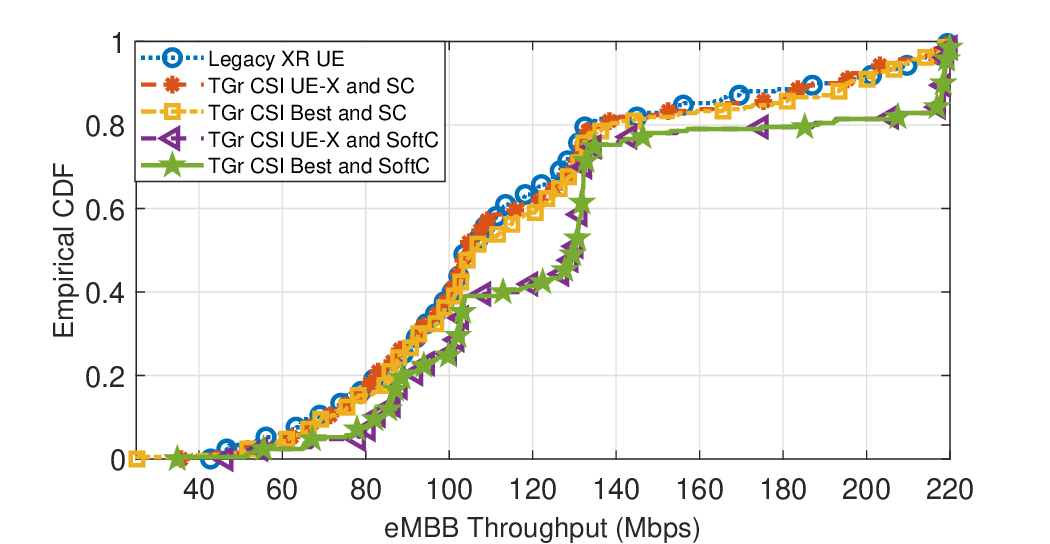}
         \caption{\gls{du}}
         \label{fig:xrcdu}
     \end{subfigure}
     \hfill
        \caption{eMBB Throughput with 4 XR UEs/TGrs per cell. (a) \gls{inh} (b) \gls{du}.}
        \label{fig:embbthroughput}
\end{figure}

\gls{xr} \gls{thgr} not only provide performance gains for \gls{xr} \gls{ue}s by increasing the \gls{xr} capacity but it also contribute to better \gls{embb} performance in the coexistence scenarios.
\gls{xr} \gls{thgr} especially with \gls{sscs} require fewer \gls{prb}s for \gls{xr} traffic, leaving more resources available for \gls{embb} traffic.
Figure \ref{fig:embbthroughput} presents the \gls{ecdf} of average \gls{embb} throughput for the cases when there is one \gls{embb} \gls{ue} with 4 legacy \gls{xr} \gls{ue}s and one \gls{embb} \gls{ue} with 4 \gls{xr} \gls{thgr}s per cell. 
The results show that \gls{xr} \gls{thgr} with \gls{sscs} achieve higher average \gls{embb} throughput than legacy \gls{xr} \gls{ue}s in both \gls{inh} and \gls{du}.
This improvement is primarily due to the reduced \gls{prb}s utilization of \gls{xr} \gls{thgr}s compared to legacy \gls{xr} \gls{ue}s for approximately similar amount of transmitted data.
The gains of \gls{xr} \gls{thgr} with \gls{scs} are smaller particularly in \gls{du}, where they are almost negligible.
Quantitatively, \gls{xr} \gls{thgr} with \gls{sscs} achieves a 43\% throughput gain for \gls{embb} at the 50\textsuperscript{th} percentile in \gls{inh} and of 25\% in \gls{du} compared to legacy \gls{xr} \gls{ue}s, whereas \gls{xr} \gls{thgr} with \gls{scs} provide only 10\% and 1\% gain in \gls{inh} and \gls{du}, respectively. 

\subsection{\gls{cb} based Cooperation}

\begin{figure}
\centerline{\includegraphics[width=8.5cm]{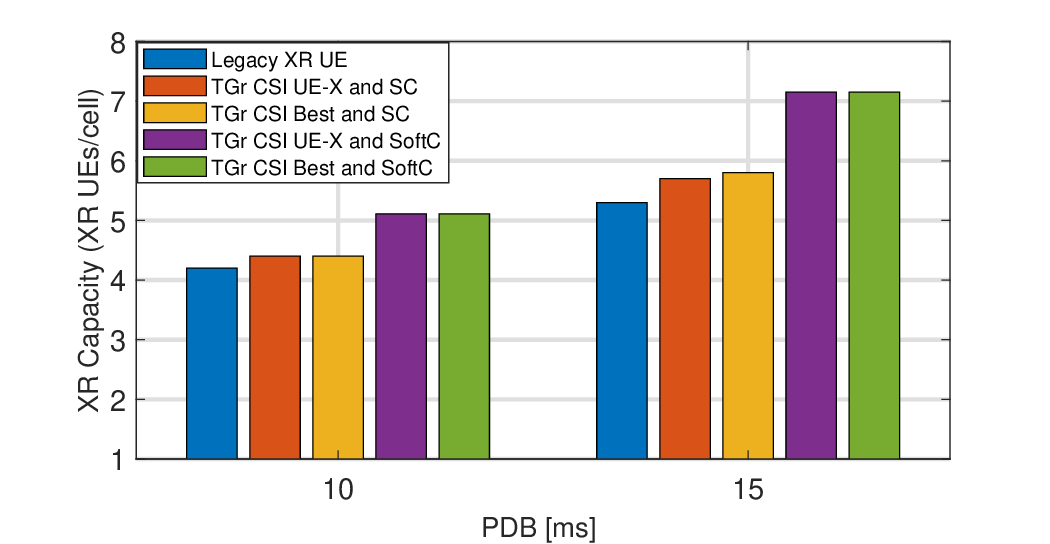}}
\caption{XR Capacity with CB based HARQ and cooperation in InH with XR UEs only.}
\label{fig:xrcapcb}
\end{figure}

This subsection summarizes the findings from the \gls{cb}-based cooperation analysis, which was conducted exclusively for the \gls{inh} deployment scenario.
Figure \ref{fig:xrcapcb} presents the \gls{xr} capacity of the \gls{inh} scenario for legacy \gls{xr} \gls{ue}s and \gls{xr} \gls{thgr} under both cooperation schemes, considering a network with \gls{xr} \gls{ue}s only.
The use of \gls{cbg} based \gls{harq} retransmissions and higher target \gls{bler} improve the performance of legacy \gls{xr} \gls{ue}s and \gls{xr} \gls{thgr} with \gls{scs} marginally compared to \gls{tb} based cooperation and \gls{harq} retransmission cases of legacy \gls{xr} \gls{ue}s and \gls{xr} \gls{thgr} with \gls{scs}, respectively.
However, the performance of \gls{thgr} with \gls{sscs} remains nearly the same as in the \gls{tb} based cooperation case.
This is because \gls{sscs} operate at maximum available \gls{mcs} most of the time with a lower \gls{bler} under the given conditions, resulting in very few \gls{harq} retransmissions, so changing the granularity of \acrshort{harq} retransmission does not affect its performance.
Nevertheless, \gls{sscs} still delivers substantial gains, ranging from 22-35\% over legacy \gls{xr} \gls{ue}s.

\begin{figure}
\centerline{\includegraphics[width=8.5cm]{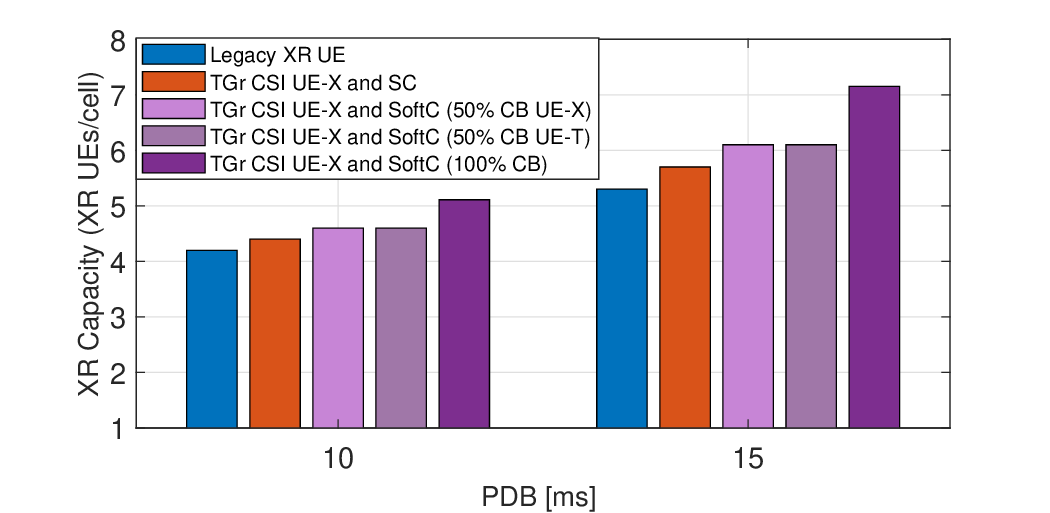}}
\caption{XR Capacity with 50\% CB in InH with XR UEs only.}
\label{fig:xrcap50cb}
\end{figure}

Figure \ref{fig:xrcap50cb} compares the \gls{xr} capacity for \gls{xr} \gls{thgr} using limited \gls{cb}-based \gls{sscs} with legacy \gls{xr} \gls{ue}s and \gls{thgr}s with \gls{sscs} using 100\% \gls{cb} in the \gls{inh} scenario. 
The results indicate that limited \gls{cb}-based cooperation achieve \gls{xr} capacity gains that lie between \gls{xr} capacity achievable through legacy \gls{xr} \gls{ue}s and \gls{xr} \gls{thgr} with \gls{sscs} using 100\% \gls{cb}s. 
The two variations of limited \gls{cb}-based \gls{sscs} exhibit almost identical average performance.
Importantly, limited \gls{cb}-based cooperation offers a practical trade-off between the amount of data exchanged over the tethering link and the performance gains achieved by the \gls{thgr}.
In the extreme case where no data is exchanged over the tethering link, the \gls{thgr}'s performance matches that of legacy \gls{xr} \gls{ue}s.

\subsection{Discussion}
The \gls{xr} \gls{thgr} with \gls{sscs} substantially improves network \gls{xr} capacity compared to legacy \gls{xr} \gls{ue}s (single link connected devices) in both \gls{inh} and \gls{du} scenarios.
By enabling \gls{xr} devices to operate at a higher \gls{mcs} while maintaining a post-cooperation \gls{bler} equal to or less than the target \gls{bler}, \gls{xr} \gls{thgr} not only reduces application-layer delay but also frees up network resources (fewer \gls{prb}s utilization).
These additional resources can be utilized to accommodate more \gls{xr} devices or to support more \gls{embb} traffic in coexistence scenarios.

The performance gains of \gls{xr} \gls{thgr} in terms of \gls{xr} capacity are nearly identical and in some cases almost equivalent, whether both \gls{ue}s in the \gls{thgr} provide \gls{csi} feedback or only one \gls{ue} per \gls{thgr} does.
This observation enables the possibility of configuring only one \gls{ue} per \gls{thgr} for \gls{csi} reporting, thereby reducing the \gls{ul} resources for \gls{csi} to the same level as in legacy \gls{xr} \gls{ue}s.
The only additional \gls{ul} resources used by \gls{thgr}, compared to legacy \gls{xr} \gls{ue}, for optimizing \gls{dl} transmissions are those required for \gls{harq} feedbacks from the tethering \gls{ue} and from the \gls{xr} \gls{ue} in \gls{sscs} after soft combining.

It is important to note that this study assumes an ideal tethering link; hence, reported network capacity improvements be regarded as upper-bound values.
In practice, the tethering link's performance will influence the realized gains and this performance will depend on the specific technology used for tethering.
For example, a tethering link operating on an independent carrier frequency than cellular network's one may introduce no additional interference to the direct link (\gls{gNodeB}-\gls{ue}).
However, latency on the tethering link could degrade \gls{thgr} performance in certain cases.
For instance, when a \gls{tb} fails on \gls{xr} device initial decoding (before any cooperation) and is already near the end of its \gls{pdb}.
In such situations, if the tethering \gls{ue} provides the correct \gls{tb} (or soft bit corresponding to that \gls{tb}) shortly afterward, the additional delay could cause a \gls{pdb} violation.
However, in such situations even \gls{harq} retransmission are insufficient.
Cooperation over the tethering link may be more effective than \gls{harq} retransmissions when the failed \gls{tb} has a margin of only a few milliseconds before \gls{pdb} expiry.
This is because the tethering link is expected to have lower latency than \gls{harq} retransmission from the \gls{gNodeB} for two key reasons, i.e., the short physical distance between the two \gls{ue}s in the \gls{thgr} and the absence of scheduling delays on the tethering link, as it could be dedicated.
In contrast, \gls{harq} retransmissions from the \gls{gNodeB} usually involve longer distances and small but non-negligible, scheduling delays, which depend on the \gls{tdd} frame structure.

\subsubsection{Complexity and Overhead Analysis of \gls{thgr}}
In this section, we discuss the complexity introduced by the \gls{thgr} for both \gls{gNodeB} and \gls{ue}s along with the signaling overhead.
\begin{itemize}
    \item \gls{gNodeB} Computational Complexity: When \gls{xr} \gls{thgr}s are enabled, the \gls{gNodeB} must collect \gls{harq} feedback from both member \gls{ue}s, take a retransmission decision based on joint \gls{harq} feedback processing and apply the enhanced joint \gls{olla} algorithm. These additional steps add only a negligible constant offset in \gls{gNodeB} complexity.
    Consequently, the additional control-plane work is a small constant per \gls{thgr} and does not change the asymptotic scaling, i.e., the \gls{gNodeB} processing remains $\mathcal{O}(N)$ for $N$ users, with only a modest constant increase.
    \item \gls{ue} Computational Complexity: In the legacy \gls{xr} \gls{ue}s case, \gls{xr} \gls{ue} performs standard demodulation/decoding procedure and return \gls{harq} feedback. In the \gls{xr} \gls{thgr} case, both \acrshort{ue-t} and \acrshort{ue-x} run the standard decoding chain and provide \gls{harq} feedback plus some additional steps depending upon the cooperation scheme. 
    \begin{itemize}
        \item \gls{scs}: \acrshort{ue-t} runs \gls{df} algorithm and forwards decoded \gls{tb} when successfully decoded. \acrshort{ue-x} performs \gls{sc} (i.e., it uses the first successful copy of the \gls{tb}). These additions impose only negligible extra arithmetic processing beyond the baseline decode operations. 
        \item \gls{sscs}: If \acrshort{ue-t} failed to decode the \gls{pdsch} but successfully decoded the \gls{pdcch} then it will forward the soft bits (\gls{llr}s). \acrshort{ue-x} then optionally performs element wise soft combining of the soft bits (\gls{llr}s) (an $\mathcal{O}(L)$ operation for a \gls{tb} with $L$ \gls{llr} entries) followed by a secondary decode and \gls{harq} feedback. Thus \gls{sscs} introduces additional per \gls{tb} arithmetic processing step that scales with \gls{tb} size.  
    \end{itemize} From an energy perspective, \gls{xr} \gls{thgr} operation can be substantially more demanding than legacy \gls{xr} \gls{ue}s because two \gls{ue}s may be decoding and one may be forwarding data. The total per \gls{thgr} energy exceeds the combined energy of two legacy \gls{xr} \gls{ue}s (due to forwarding of the data and selection or soft combining).  
    \item Signaling Overhead: A legacy \gls{xr} \gls{ue}s provide one \gls{harq} feedback per \gls{tb} (usually one bit if \gls{harq} feedback is per \gls{tb}) for the initial transmission and one \gls{csi} report (periodic or aperiodic). For \gls{xr} \gls{thgr}:
    \begin{itemize}
        \item \gls{scs}: two \gls{harq} feedback per \gls{tb} for the initial transmission.
        \item \gls{sscs}: two initial \gls{harq} feedback per \gls{tb} plus a conditional secondary \gls{harq} feedback from \acrshort{ue-x} only when soft combining is performed.
    \end{itemize}
    To limit resources utilized for \gls{csi}, the \gls{xr} \gls{thgr} could be configured such that only one \gls{ue} per \gls{thgr} report \gls{csi} with negligible performance degradation in many scenarios, so resources utilized for \gls{csi} would be similar for both legacy \gls{xr} \gls{ue}s and \gls{xr} \gls{thgr}s. Since \gls{harq} feedback is extremely small often just one bit for \gls{tb} based \gls{harq} retransmissions, so the signaling overhead (\gls{ul} resource cost) of enabling these \gls{dl} gains is very small. Moreover, in lower user densities, the cooperation gain of the \gls{xr} \gls{thgr} cannot choose higher \gls{mcs} then the legacy \gls{xr} \gls{ue}s (maximum \gls{mcs} is capped and legacy \gls{xr} \gls{ue}s are already operating at maximum \gls{mcs}) but this cooperation reduces the number of retransmissions from the \gls{gNodeB} which tends to lower overall \gls{gNodeB} scheduling and physical layer load for the given user.  
\end{itemize}

Enabling \gls{xr} \gls{thgr} requires only small changes to the standard, slight modifications in \gls{gNodeB} control logic with small constant overhead and introduces modest additional \gls{ue} computation. Taken together, the complexities and overheads are small compared with the substantial gains in spectral efficiency that this cooperative framework delivers especially for the time-bound \gls{xr} application. 

\section{Conclusion}

\label{conc}
This paper presented an approach for optimizing \gls{ptm} transmission to multi-connected \gls{xr} \gls{thgr} with the objective of enhancing network \gls{xr} capacity.
An enhanced joint \gls{olla} algorithm was proposed to exploit the benefits of either \gls{scs} or \gls{sscs} cooperation scheme in the \gls{la} process for \gls{ptm} transmission.
The method leverages \gls{harq} feedback from both \gls{ue}s in an \gls{xr} \gls{thgr} to optimize the \gls{mcs} based on the post-cooperation \gls{bler} of the \acrshort{ue-x}.
Coupled with a joint \gls{harq} feedback processing algorithm, the proposed enhanced joint \gls{olla} enables the network to operate at higher spectral efficiency, thereby reducing \gls{prb} usage per \gls{xr} user and lowering application layer delay.
The delay reduction and lower \gls{prb} utilization increases the likelihood of meeting the \gls{xr} "happiness" criteria for more users, while extra \gls{prb}s can also be utilized to support more \gls{embb} traffic, ultimately boosting overall network capacity.
The algorithm also demonstrates robustness in \gls{mcs} optimization, achieving near identical performance even when only one \gls{ue} per \gls{thgr} is configured to provide \gls{csi}, thus reducing the \gls{ul} resource overhead for \gls{xr} \gls{thgr} operation under the given channel conditions.

Dynamic system level simulations show that \gls{xr} \gls{thgr}s with joint \gls{harq} feedback processing and enhanced joint \gls{olla} achieve \gls{xr} capacity gains in the range of 23-42\% with \gls{sscs} cooperation in \gls{xr}-only scenarios.
These gains become even more pronounced under higher-interference coexistence scenarios involving \gls{xr} and \gls{embb} users, where \gls{sscs} cooperation yields improvements in the range of 38-173\%.
Furthermore, in coexistence scenarios with four \gls{xr} users and one \gls{embb} user, \gls{embb} throughput increases by 25-43\% due to lower \gls{prb} utilization by \gls{xr} \gls{thgr}s.

\section*{Abbreviations}
\label{appen}
\begin{tabbing}
    \hspace{2cm} \= \hspace{4cm} \= \kill

    3GPP \> 3rd Generation Partnership Project\\
    5G-A \> 5th generation-advanced \\

    ACK \> Acknowledgment \\
    AR \> Augmented reality\\
    BLER \> block error rate\\
    CB \> Code block\\
    CBG \> Code block group\\
    CC \> Chase combining\\
    CSI \> Channel state information\\
    CQI \> Channel quality indicator\\
    DU \> Dense urban \\
    DL \> Downlink\\
    DF \> Decode-and-forward\\
    eMBB \> Enhanced mobile broadband\\
    eCDF \> Empirical cumulative distribution function\\
    gNodeB \> Next-generation node B\\
    HARQ \> Hybrid automatic repeat request\\
    ILLA \> Inner loop link adaptation\\
    InH \> Indoor hotspot\\
    ICI \> Inter-cell interference\\
    LA \> Link adaptation \\
    LOS \> Line-of-sight\\
    LLR \> Log-likelihood ratio\\
    MMIB \> Mean mutual information per bit\\
    mmWave \> Millimeter-wave\\
    MIMO \> Multiple-Input and Multiple-Output\\
    MBS \> Multicast and broadcast services\\
    MMSE-IRC \> Minimum mean square error-interference \\
    \> rejection combining \\
    MCS \> Modulation and coding scheme\\
    MR \> Mixed reality\\
    NR \> New radio\\
    NACK \> Negative acknowledgment\\
    NLOS \> Non-line-of-sight\\
    OFDM \> Orthogonal frequency-division multiplexing\\
    OLLA \> Outer loop link adaptation\\
    PDB \> Packet delay budget\\
    PTM \> Point-to-multipoint\\
    PTP \> Point-to-point\\
    PRB \> Physical resource block\\
    PDSCH \> Physical downlink shared channel\\
    PDCCH \> Physical downlink control channel\\
    QoS \> Quality of service\\
    RRM \> Radio resource management\\
    Rx \> Receive\\
    SC \> Selection combining\\
    SCS \> Selection combining Scheme\\
    SSCS \> Selection/Soft combining scheme\\
    SVD \> Singular value decomposition\\
    soft-DF \> Soft decode-and-forward\\
    SINR \> Signal-to-interference-plus-noise ratio\\
    SLS \> System-level simulator\\
    TBLER \> Transport block error rate\\
    TGr \> Tethering group\\
    TB \> Transport block\\
    Tx \> Transmit\\
    TDD \> Time division duplex\\
    UE \> User equipment\\
    UE-X \> User equipment-extended reality\\
    UE-T \> User equipment with tethering\\
    UL \> Uplink\\
    URLLC \> Ultra reliable low latency communications\\
    VR \> Virtual reality\\
    WLAN \> Wireless local-area network\\
    XR \> Extended reality\\

\end{tabbing}

\begin{IEEEbiography}[{\includegraphics[width=1in,height=1.25in,clip,keepaspectratio]{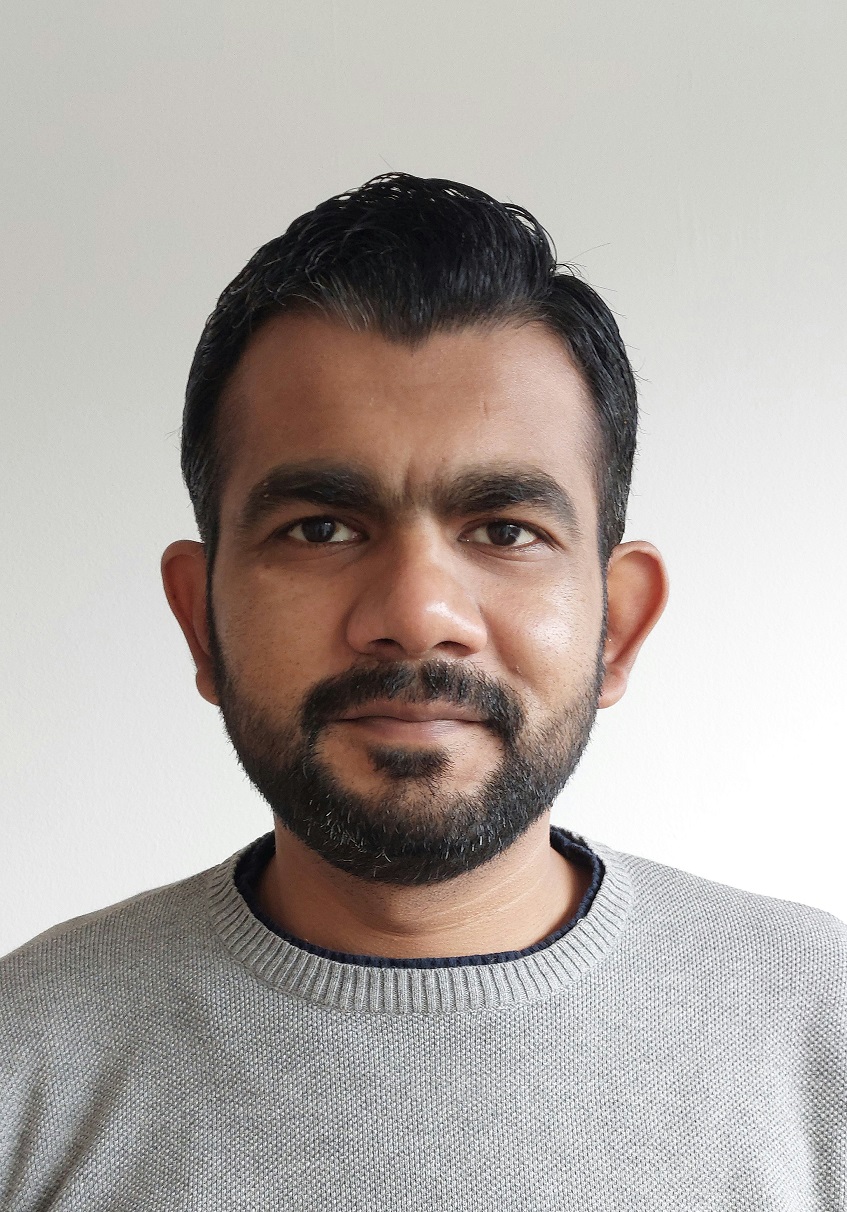}}]{\textbf{MUHAMMAD AHSEN }}  (Graduate Student Member, IEEE) received the B.E. degree in Electrical Engineering in 2013 and the M.S. degree(Hons.) in Electrical Engineering in 2016 from the National University of Sciences and Technology (NUST), Islamabad, Pakistan. He is currently pursuing the Ph.D. degree with the Department of Electronics Systems, Aalborg University, Denmark, in collaboration with Nokia Standard, Aalborg, Denmark. His research interest includes extended reality communications, 5G/6G radio resource management, cooperative communication and time-sensitive wireless communication.  
\end{IEEEbiography}

\begin{IEEEbiography}[{\includegraphics[width=1in,height=1.25in,clip,keepaspectratio]{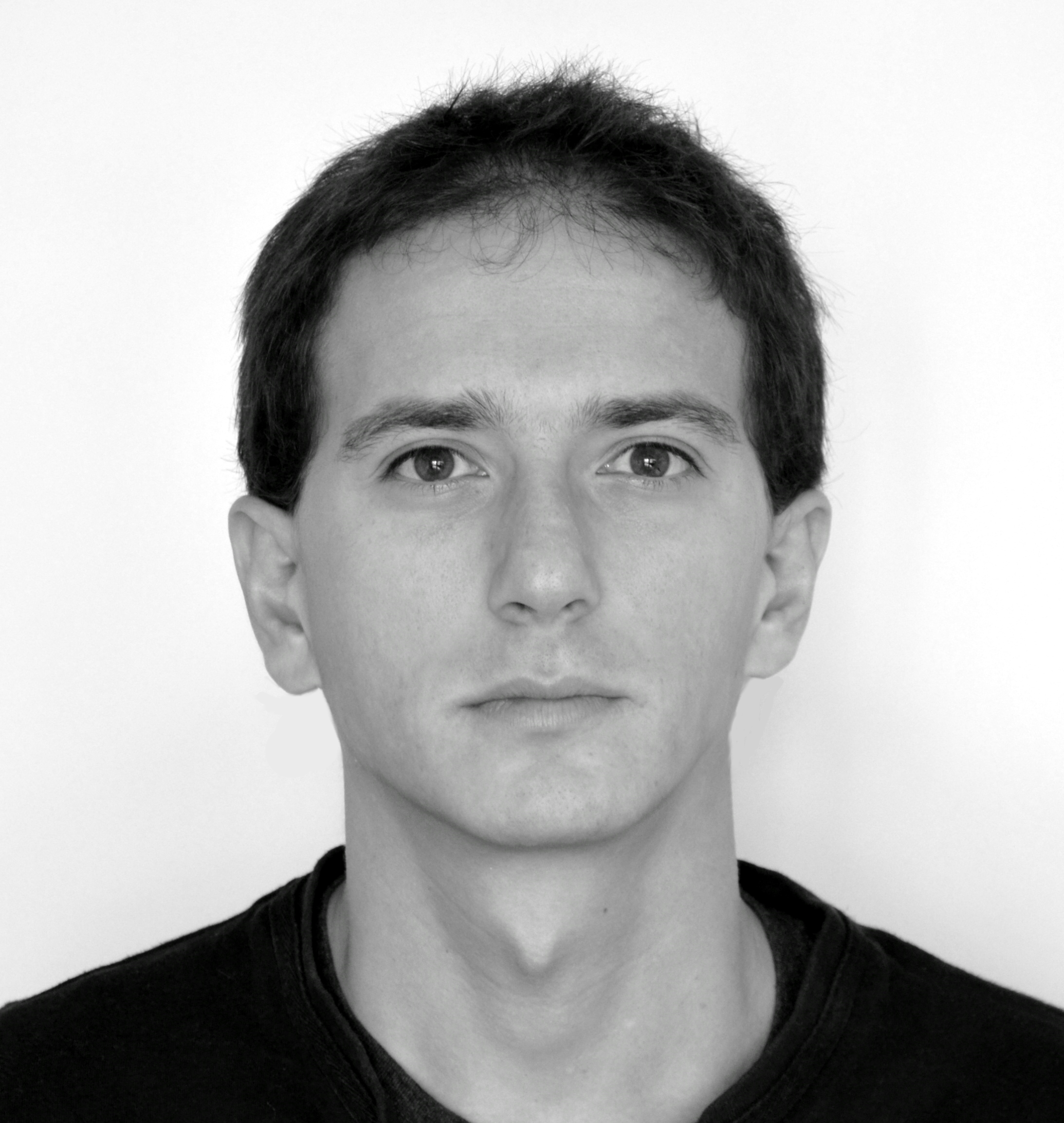}}]{\textbf{BOYAN YANAKIEV }} received the B.Sc. degree in physics from Sofia University, Bulgaria, in 2006, and the M.Sc. and Ph.D. degrees from Aalborg University, Aalborg, Denmark, in 2008 and 2011, respectively. He is currently with Nokia Standardization Research, Denmark, focusing on 5G advanced and 6G topics. His current work is in the area of XR optimization in RAN  
\end{IEEEbiography}

\begin{IEEEbiography}[{\includegraphics[width=1in,height=1.25in,clip,keepaspectratio]{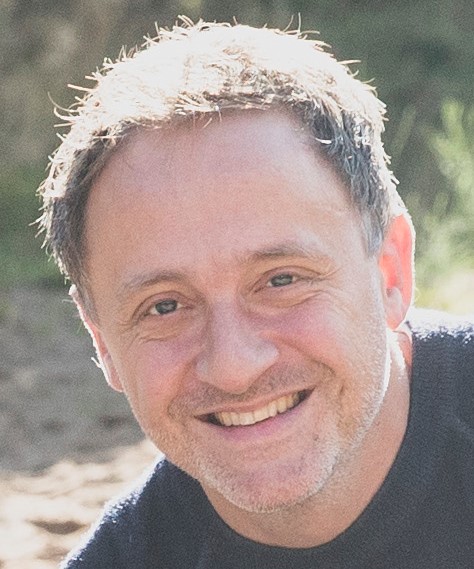}}]{\textbf{CLAUDIO ROSA }} received his M.Sc. E.E. and Ph.D. degrees in 2000 and 2005, respectively, from Aalborg University. In 2003 he also received an M.Sc.E.E. degree in telecommunication engineering from Politecnico di Milano, Italy. Since he joined Nokia in 2005, he contributed to standardization of 4G and 5G systems working on uplink power control and radio resource management, carrier aggregation, dual connectivity, and unlicensed spectrum operation. His current research interests also include user plane protocol design for 6G, flexible duplexing and radio enablers for eXtended reality applications. He has filed more than 200 patent applications, holds more than 50 granted patents, and has co-authored more than 50 scientific publications.   
\end{IEEEbiography}

\begin{IEEEbiography}[{\includegraphics[width=1in,height=1.2in,clip]{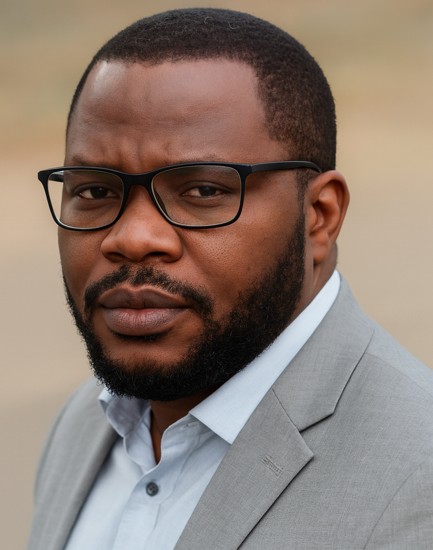}}]{\textbf{RAMONI ADEOGUN }} (Senior Member, IEEE) received the B.Eng. degree in electrical and computer engineering from the Federal University of Technology, Minna, Nigeria, and the Ph.D. degree in electronic and computer systems engineering from the Victoria University of Wellington, New Zealand in 2007 and 2015, respectively. He is currently an Associate Professor and leader of the AI for Communications group at Aalborg University, Denmark. Prior to joining Aalborg University, he held varios academic and industry research positions at University of Cape Town, South Africa, Odua Telecoms Ltd., and the National Space Research and Development Agency, Nigeria. His research interests include signal processing, machine learning and AI for PHY MAC and RRM. He has co-authored over 70 peer-reviewed publications. 
\end{IEEEbiography}

\end{document}